# Critical assessment of the evidence for striped nanoparticles

Julian Stirling[1,*], Ioannis Lekkas[1], Adam Sweetman[1], Predrag Djuranovic[2], Quanmin Guo[3], Josef Granwehr[4], Raphaël Lévy[5], Philip Moriarty[1]

**1 School of Physics and Astronomy, The University of Nottingham, University Park, Nottingham, NG7 2RD, United Kingdom**
**2 Department of Materials Science and Engineering, Massachusetts Institute of Technology, 77 Massachusetts Avenue, Cambridge, MA 02139, USA**
**3 School of Physics and Astronomy, University of Birmingham, Birmingham B15 2TT, United Kingdom**
**4 Sir Peter Mansfield Magnetic Resonance Centre, School of Physics and Astronomy, The University of Nottingham, University Park, Nottingham, NG7 2RD, United Kingdom**
**5 Institute of Integrative Biology, Biosciences Building, University of Liverpool, Crown Street, Liverpool, L69 7ZB, United Kingdom**
**∗ E-mail: ppxjs1@nottingham.ac.uk**

## Abstract

There is now a significant body of literature in which it is claimed that stripes form in the ligand shell of suitably functionalised Au nanoparticles. This stripe morphology has been proposed to strongly affect the physicochemical and biochemical properties of the particles. We critique the published evidence for striped nanoparticles in detail, with a particular focus on the interpretation of scanning tunnelling microscopy (STM) data (as this is the only technique which ostensibly provides direct evidence for the presence of stripes). Through a combination of an exhaustive re-analysis of the original data with new experimental measurements of a simple control sample comprising entirely unfunctionalised particles, we conclusively show that all of the STM evidence for striped nanoparticles published to date can instead be explained by a combination of well-known instrumental artefacts, strong observer bias, and/or improper data acquisition/analysis protocols. We also critically re-examine the evidence for the presence of ligand stripes which has been claimed to have been found from transmission electron microscopy, nuclear magnetic resonance spectroscopy, small angle neutron scattering experiments, and computer simulations. Although these data can indeed be interpreted in terms of stripe formation, we show that, just as for the STM measurements, rather more mundane interpretations can account for the reported results.

## Introduction

Scanning probe microscopy (SPM) is an exceptionally powerful technique at the core of modern nanoscience. Indeed, many would argue that the origins of the entire field of nanoscale science lie in the invention of the scanning tunnelling microscope (STM) in the early eighties [1]. Single atoms and molecules are now not only routinely resolved with STM but, under appropriate experimental conditions, can be precisely positioned [2–5] to form artificial nanostructures exhibiting fascinating quantum mechanical properties [6–8].

The development of the atomic force microscope (AFM) [9] shortly after the introduction of the STM broadened the applicability of SPM to a much wider variety of substrates — including, in particular, insulators — and led to the adoption of SPM as a high resolution imaging technique in very many scientific disciplines and sub-fields. The state of the art in atomic force microscopy is no longer 'just' atomic resolution [10] (a remarkable achievement in itself), but the imaging of intramolecular [11–13] and intermolecular *bonds* [14, 15]. Furthermore, SPM systems now operate in a range of environments spanning what might be termed 'extreme' conditions — ultrahigh vacuum, low temperatures, and high magnetic fields (for example, an STM running at 10 milliKelvin in a field of 15 T has recently been developed [16]) — to the *in vitro* application of AFM to study biochemical and biomedical processes [17].

A significant number of commercial suppliers also now provide 'turn-key' SPM systems such that the probe microscope has evolved into a standard characterisation tool in the vast majority of nanoscience laboratories.

Unfortunately, however, with the exceptional capabilities of the scanning probe microscope come a plethora of frustrating instrumental artefacts. These can give rise to images which, although initially appearing entirely plausible, unsettlingly arise from a variety of sources including improper settings of the microscope parameters (for example, the feedback loop gains used to control the motion of the scanning probe), external electrical or vibrational noise, and/or convolution of the sample topography with the structure of the probe. The latter is especially problematic when the features of interest at the sample surface have a radius of curvature which is comparable to that of the tip.

While some of these SPM artefacts, such as those due to improper feedback loop settings, are relatively straight-forward to diagnose and eliminate, tip-sample convolution can often require particularly careful and systematic experimental technique to identify and remove [18]. Debates in the literature regarding artefacts in atomic/molecular resolution images arising from, e.g., 'double' or multiple tips [19], and/or tip asymmetry [20], show that, unless appropriate experimental protocols have been used to ensure that the SPM images are as free of tip influence as possible, it can be exceptionally difficult to deconvolve the influence of the tip structure from the final image. In addition, without appropriate control samples it is entirely possible to misinterpret genuine and mundane surface features as new and hitherto unobserved aspects of the molecule or structure of interest. This latter problem was brought sharply to the fore in the early days of STM when the results of very high profile papers claiming to have attained high resolution images of DNA and other biomolecules on graphite were replicated on freshly cleaved, i.e. entirely molecule-free, substrates. The 'molecular' images were shown in a number of cases to arise from step edges and graphitic fragments ("flakes") on the bare graphite surface [21].

In this paper we critique, in the context of the SPM artefacts described above, the body of highly-cited work published by Stellacci and co-workers over the last decade or so [22–26], which claims that stripes form in the ligand shell of appropriately functionalised gold nanoparticles. These claims have subsequently led to the proposal that ligand stripes substantially influence the ability of nanoparticles to penetrate cell membranes [25], and, very recently, Cho *et al.* [26] have argued that the striped morphology enables high selectivity for heavy metal cations (although there are unresolved issues regarding the lack of appropriate control samples for this study [27]). By combining an extensive re-analysis of Stellacci *et al.*'s data with imaging of a simple control sample comprising ligand-free nanoparticles, we show that the scanning probe data published to date provide no evidence for stripe formation and instead can be explained by a combination of instrumental artefacts, data selection ('cherry picking'), and observer bias. For completeness, we also consider the evidence, or lack thereof, for stripe formation from other techniques such as transmission electron microscopy [23], nuclear magnetic resonance (NMR) spectroscopy [28], and computer simulations [29]. Taken together, our analyses provide important insights into the pitfalls of not adopting an extremely critical, systematic, and sceptical approach to SPM imaging of nanostructured samples.

## Materials and Methods

In order to demonstrate how striped features and other intraparticle structure can arise from STM artefacts, we prepared a control sample comprising entirely unfunctionalised nanoparticles. This was generated under ultrahigh vacuum conditions so as to ensure that the nanoparticle surfaces remained free of contamination and adsorbates.

Following a well-established approach [30,31], a $C_{60}$ monolayer (ML) was formed on the Si(111)-(7x7) surface to act as a template for the formation of Ag nanoparticles. $C_{60}$ was first sublimated onto a clean Si(111)-(7x7) surface, which had been formed using standard flash annealing procedures [32]. Following the deposition of a multilayer fullerene film, the sample was annealed at $\sim 450°C$ to desorb all $C_{60}$ other than the first chemisorbed monolayer. Ag was then deposited from a Knudsen cell operating at a temperature of approximately 880°C onto the 1 ML $C_{60}$/Si(111) sample. In order to modify the size distribution of the Ag nanoparticles — so as to make the particles' mean diameter comparable to that of those studied by Stellacci *et al.* — we subsequently annealed the Ag-covered $C_{60}$ monolayer sample in the 200°C to 400°C range.

Our STM measurements were acquired using an Omicron Nanotechnology low temperature ultrahigh vacuum qPlus atomic force microscope–scanning tunnelling microscope instrument operating at 77 K at a pressure of $\sim 5 \times 10^{-11}$ mbar. All SPM image analysis in this paper is performed using scripts written in MATLAB using the SPIW toolbox [33]. The raw data and scripts have been made public [34] to allow our analysis to be repeated and/or modified by any interested party.

## Results and Discussion

In the following sections we re-analyse the evidence for striped nanoparticles that has been presented by Stellacci and co-workers in a series of papers over the last decade. Where necessary, we complement the re-analysis of Stellacci *et al.*'s data with a discussion of STM measurements of the Ag nanoparticle sample described in the preceding section. A key advantage of the protocol we have adopted for nanoparticle synthesis is that the Ag particle surfaces in our experimental measurements are entirely ligand free. As such, they act as excellent control samples to highlight the role of instrumental artefacts and improper data acquisition/analysis protocols when making claims for structure in a ligand shell.

Following criticism of the evidence for stripes by Cesbron *et al.* [35], some raw STM data from the first papers published by Stellacci *et al.* [23, 24, 36] was placed in the public domain [37]. For reasons detailed in the following sections, the archived data do not, however, justify the conclusions drawn in these papers. A number of other papers based on STM data have also been published since the archived data was released [38–40] and we are grateful to the corresponding author of one of those papers [39] for providing some of the data associated with that work for re-analysis. We examine and provide a detailed critique of this STM data, and we discuss the evidence, or lack thereof, for stripe formation from a variety of other techniques.

### Striped features in SPM images arising from feedback instabilities

The first paper on the striped morphology of Au nanoparticles (Jackson *et al.* 2004 [23]), leads with an STM image of nanoparticles having a mixed 1-octanethiol (OT) and mercaptopropionic acid (MPA) termination, showing striped features on each nanoparticle (reproduced in Figure 1c below). This is one of the clearest STM images of stripes of which we are aware and played a seminal role in establishing the concept of "striped" ligand patterns on Au nanoparticles. Before we discuss the compelling evidence that the stripes simply arise from a well-known STM artefact, and not from ligand organisation, we first note that the contrast in the image is saturated at the lower end of the contrast scale (i.e black). If we instead set a linear contrast scale from the highest to lowest pixel (as is standard practice) it is clear that *the stripes extend between the nanoparticles* (Figure 1b). This observation alone strongly suggests that the stripes are not real surface features confined to the nanoparticles. We note in passing that the image from Jackson *et al.* 2004 [23] included as Figure 1c is a $38 \times 38$ $nm^2$ offline zoom of a $157 \times 157$ $nm^2$ image (Figure 1a). To increase the apparent resolution the image has then been interpolated up to a much larger number of pixels, and possibly filtered to give rounded shapes to features which are only 2–3 pixels across. We will return to a discussion of how this type of image processing can give rise to misleading results.

To understand scanning probe microscopy image artefacts it is first necessary to realise that the images are formed by bringing a sharp tip close to the surface under study. In the case of STM, a feedback loop controlling the tip-sample separation is used to maintain a constant tunnelling current between the tip and the surface. By recording the 3D path taken by the tip as it is raster scanned over the surface, a height profile is taken. Improper choice of scan speed or feedback gains can result in poor regulation of tunnel current or even complicated feedback instabilities. In addition, as the current to be regulated is of the order of nanoamps, the effect of electrical noise cannot be neglected. Furthermore, even assuming perfect feedback conditions, the image is a convolution of the surface and tip structure, combined with the presence of tip-sample forces, which can cause changes to either (or both) during the image acquisition, resulting in abrupt modifications.

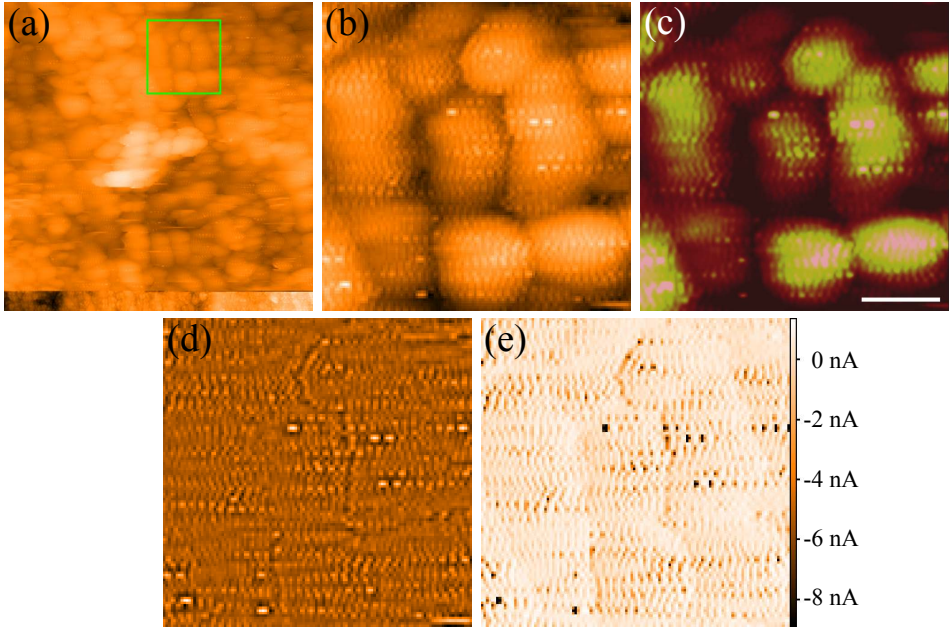

**Figure 1. Reanalysis of the data for Figure 1a) of Jackson** *et al.* **2004 [23]. (a)** The raw 157 nm wide image collected for Jackson *et al.* 2004. **(b)** Zoom-in on a 37 nm wide area marked in green on (a). This image has been flattened using first order plane subtraction. It is clear that the ripples extend between the particles. **(c)** Figure 1a) from Jackson *et al.* 2004. Note that the choice of contrast obscures the ripples between the particles. Scale bar 10nm. **(d)** Fourier transform high-pass filter of (b), removing spatial frequencies below $0.33 \times 10^9$ m$^{-1}$. **(e)** Simultaneous current image of (b). Note the (inverted) similarity to (d). The colour ranges for (a), (b) and (d) are set to run linearly from the highest to the lowest pixel. For (e) the colour range is set to run linearly for the centre 99.6% of pixels, as extreme pixels mask much of the contrast (For this section of the image the tunnel current spans a range from -51.2 nA to 2.83 nA). Colour bar shows recorded current values, the setpoint current is +838 pA.

Thus, to reliably verify the existence of specific topographic structure it is important to systematically probe the features by comparing the trace and retrace images from the STM, taking repeat scans of the same feature, rotating the scan direction, deliberately modifying the tip in order to ascertain the level of tip-sample convolution, and zooming in on specific features in 'real time', i.e. by reducing the scan area imaged by the STM, to check that features are unchanged [41]. We stress that in the majority of Stellacci *et al.*'s work, and certainly for the original, highly cited Jackson *et al.* 2004 [23] paper, these basic checks on image consistency have not been carried out.

To help the user identify artefacts arising from improper feedback settings, scanning probe microscopes normally also save images from other data channels in addition to the topography channel. An important diagnostic tool is the error signal (or current image), i.e. the difference between the setpoint value and the measured current. Ideally, the current image should be blank, but as the feedback is not instantaneous there is normally some surface structure visible. Strong, clear features in the tunnel current image, however, imply the feedback is not performing correctly. More importantly, the values of the pixels in the tunnel current image should not differ dramatically from the setpoint current used to acquire the topographic image.

Figure 1e shows the tunnel current image recorded simultaneously with Figure 1b. The structure from the topography image is clearly visible in the tunnel current map. It is possible to remove the curvature of the nanoparticles from the topography using a Fourier transform approach to filter out spatial frequencies below $0.33 \times 10^9 \mathrm{~m}^{-1}$ (Figure 1d). This further enhances the similarity to the tunnel current image, strongly suggesting that any sub-nanoparticle resolution results solely from tunnel current tracking errors. The full code used to generate Figure 1 is presented in the supplementary information.

There are, however, even more fundamental problems with the tunnel current image shown in Figure 1e). From the data archive placed in the public domain by Stellacci *et al.*, we find that the image was taken with a current setpoint of 838 pA and a sample bias of 1V (despite the text of the paper stating the images were recorded with setpoints of 500–700 pA). Pixel values from the (full) current image range from 20.2 nA to -98.2 nA. These values are clearly unphysical as the current changes sign while the voltage does not. The tunnel current values have been confirmed in Gwiddion [42], WSxM [43], and NanoScope Ver 5.31r (the software used to record the original image). It is important to note that programs such as WSxM and NanoScope automatically pre-process images by background subtracting or truncating the $z$-range. Such pre-processing must be turned off to restore the correct current values.

A possible explanation for these results, as suggested by the Stellacci group [44], is that their microscope was set to automatically background subtract the tunnel current data before saving, and thus the raw images were never correctly saved. The implications of this are that the true current range, which should be largely unaffected by background subtraction, is of order 118 nA. If we apply the most fair attempt at inverting this subtraction by shifting all pixel values until the lowest point reaches zero this would give a mean tunnel current of order 98 nA, orders of magnitude above the setpoint. This is far above the normal range of currents expected for accurate STM measurements of nanoparticle assemblies.

Another explanation is that the current-to-voltage amplifier saturates to a value of -100 nA for any currents outside its measurement range of $\pm 100$ nA. Negative pixels result from averaging of positive signals with -100 nA during saturation. This explanation would imply that the current preamp was regularly saturating to over 100 nA while feedback tries to maintain a setpoint of less than 1 nA.

The key point is that, regardless of which explanation for the negative current values is correct, the tunnel current image clearly exhibits exceptionally strong oscillations in the error signal. These arise from improper setting of the feedback loop gains (and other scan parameters). It is thus feedback loop oscillation, and not the self-assembly of two different ligand types, which gives rise to the stripes observed in the STM images shown in Jackson *et al.* 2004 [23].

Further images produced by Stellacci *et al.* show very similar contrast to those included in Jackson *et al.* 2004 [23], including, for example, Figure 2a) reproduced from Uzun *et al.* [45]. To assess whether these images also arise from feedback artefacts, and without having access to the raw data, we have used simulated SPM feedback to generate expected images from improper feedback settings, as shown in Figures 2c–i). Before giving details of the simulation we present a brief summary of how feedback is implemented in a real STM.

STM feedback utilises a proportional-integral (PI) controller feedback mechanism, similar to the common proportional-integral-derivative controller, but without the derivative component, as this acts as a high pass filter amplifying noise. The proportional part of this controller simply records the error signal (the difference between the setpoint amplitude and the recorded amplitude), multiplies this by a gain factor ($K_p$) and adds this to the extension of the piezo. The integral part of the controller integrates the error signal over time and multiplies by a separate gain ($K_i$). This removes steady state errors which arise from effects such as sample drift and cannot be corrected using simply a proportional controller. The trade off with adding the integral controller is that the tip position overshoots the optimal position before returning. If $K_i$ is too large the feedback can become unstable and oscillate about the optimal position. Therefore, for stable imaging it is necessary to carefully adjust $K_i$ and $K_p$ in order to reduce the error signal.

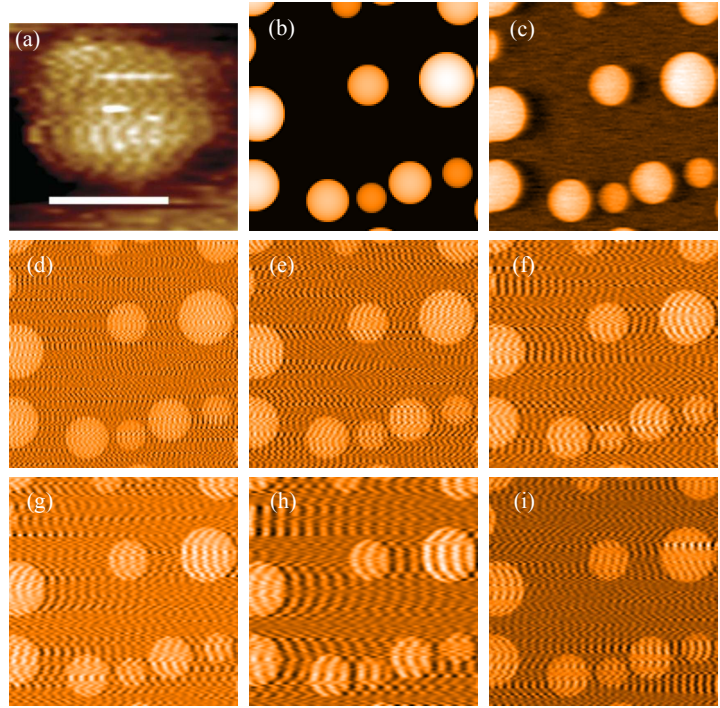

**Figure 2. Comparison of STM image of nanoparticle "stripes" with simulated STM feedback results. (a)** Image from Uzun *et al.* [45], showing features which can be reproduced by simulated SPM feedback (scale bar 5 nm). **(b)** Surface topography used in all numerical simulations. **(c)** Numerically simulated image with appropriate parameters $K_p = 500$ and $K_i = 100$. **(d–h)** The same simulation with $K_p = 50$ and $K_i = 8000, 5000, 3000, 2000,$ and $1000$ respectively. Image (i) is the retrace image recorded while recording image (f) presented directly above.

A real STM controller performs all measurements at discrete time intervals and does all calculations numerically. As such, we have written a numerical simulation, which mimics the STM's response to a given topography, by implementing a PI controller (Figure 2b)). For this simulation each measurement is subject to white noise to simulate electrical noise. Full details of the simulation, and all code used, are provided in the Supplementary Information files. Analytical methods for this type of control theory modelling are available and have very recently been used by Stellacci and co-workers [39]. We stress, however, that the method adopted by Stellacci *et al.* [39] inadvertently produces oscillations arising from incorrect modelling of mechanical components and the PID loop itself, rather than from feedback instabilities [46].

Using the simulation methods described in the Supplementary Information, it is straightforward to generate images of a smooth surface which appear to show stripes (Figure 2d–h)) by choosing an inappropriately high integral gain coupled with a low proportional gain. As the integral gain is increased the width of the stripes can be modified. Using more appropriate imaging parameters (Figure 2c)) instead, the surface structure can be accurately reproduced. It is also important to note that when the trace and retrace images — recorded when the tip is rastering in opposite directions — are compared, the curvature of the stripes changes (Figure 2f and i). This difference between trace and retrace images is a common method used to identify feedback instabilities but, unfortunately, until only very recently was not used by Stellacci *et al.*

To complement the results of the simulations we have grown Ag nanoparticles using the procedure described in Materials and Methods, and imaged these particles under various gain conditions with an Omicron low temperature STM. One minor disadvantage of using the Omicron microscope for this test is that the proportional and integral gains cannot be varied separately using the control software. Instead, a combined feedback gain is set as a percentage of the maximum allowed gain.

Figure 3 shows consecutive images of the same nanoparticle taken at increasing gains. In agreement with the simulations shown in Figure 2, at an appropriate gain setting the STM image shows the bare featureless nanoparticle surface. As the gain is increased striped features appear in the STM image. In addition, and as predicted by the simulation, the stripes vary in both contrast and width as the gain is increased. As the frequency of feedback oscillations is dependent on both the proportional and integral gain, which are not known separately for the Omicron system, we cannot directly compare the evolution of stripes in the experiment with those in the simulation (where the proportional gain is constant). All images represent a 71 pixel × 71 pixel section of a 512 pixel × 512 pixel image, which was then bicubically interpolated up to 284 pixels × 284 pixels to mimic the interpolation in published STM images of "striped" nanoparticles.

### Assessing the statistical analysis used to distinguish artefacts from real structure

Notwithstanding the discussion in the previous section, Stellacci *et al.* have argued that they can distinguish between feedback loop artefacts and true nanoparticle topography. In two publications [24, 36] following the Jackson *et al.* 2004 [23] paper critiqued above, a "statistical analysis" of previous STM data (from their group) was used to claim that feedback artefacts could be differentiated from real topographical structure. In this section we critically consider the evidence for that claim. Before doing so, it is perhaps worth noting that an experimental protocol, which involves setting abnormally high loop gains to distinguish between "real" stripes and those due to high loop gains is not a particularly robust approach to making STM measurements. A rather more compelling strategy would be to ensure that the loop gains were set appropriately and to demonstrate that, under conditions where the tip is accurately tracking the surface, stripes similar to those shown in Jackson *et al.* 2004 [23] remain visible. Throughout all of the work published by Stellacci *et al.* this has not been achieved. We return to this point repeatedly below.

The key claim of Jackson *et al.* 2006 [24] is that it is possible to distinguish between noise and ripples arising from real nanoparticle structure. In Figure 3 of that paper [24] changes in noise and ripple spacing as a function of tip speed are shown. The caption for that figure states that "*Each point in the plots is the average of multiple measurements*". This is highly misleading, however, as only one image, *of a different surface area each time*, was taken for each tip speed. The multiple "measurements" are, therefore, simply multiple readings of spacings of different features in the same image, and *not of the same particle.*

The spacings described in Jackson *et al.* 2006 [24] were determined by measuring the separation between high intensity pixels in the images — which, again, are interpolated zooms of larger area scans — and are quoted in the image annotation to a rather optimistic significance of 10 pm (It is worth noting that 10 pm equates to a separation of 0.026 pixels in the raw, uninterpolated image). The distances measured range from approximately 2 to 4 pixels and thus are very close to the (Nyquist) resolution limit of a 2 pixel spacing. We note that this combination of large area scanning followed by highly interpolated offline zooms is a rather unorthodox approach to scanning probe microscopy that, for good reason, is not widely applied within the SPM community.

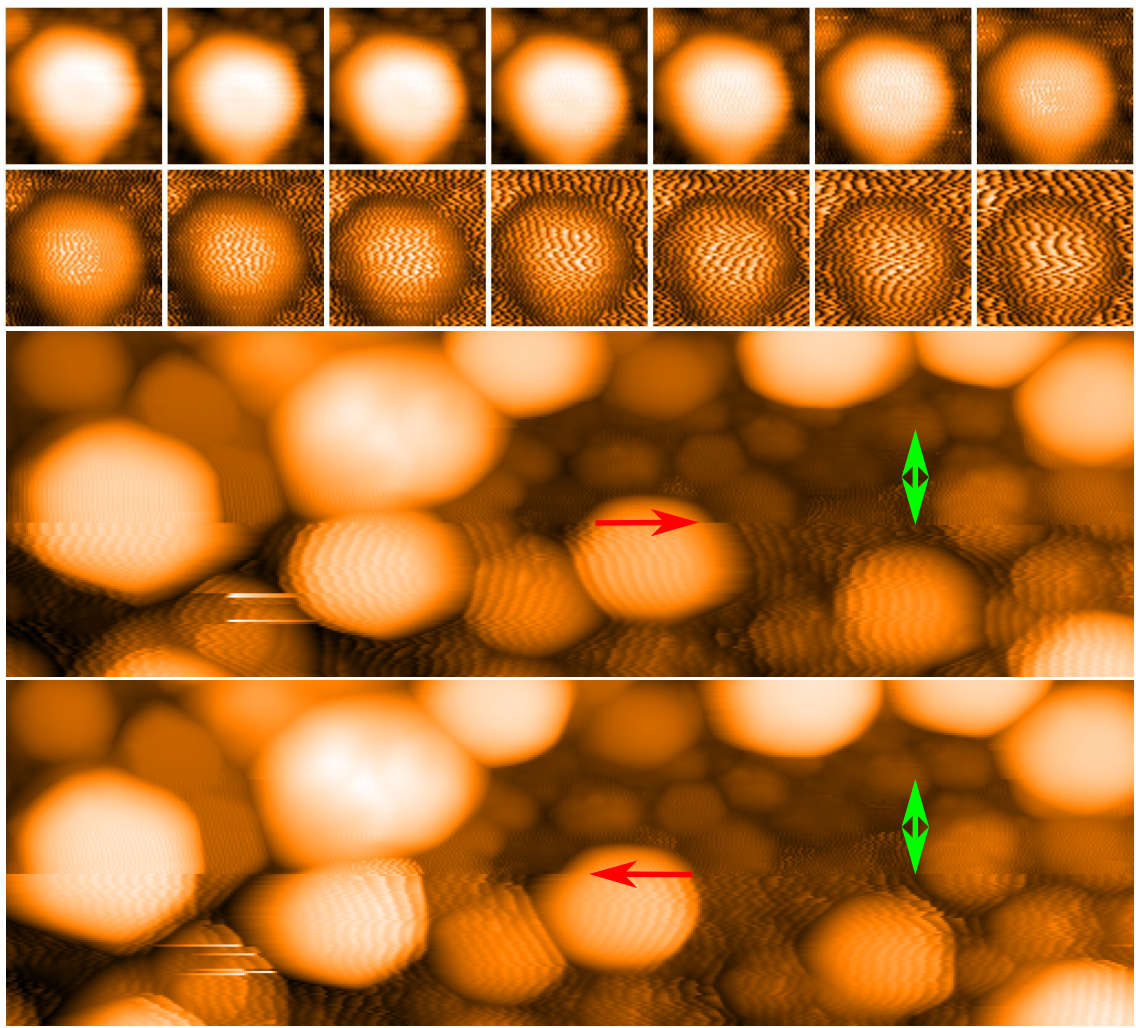

**Figure 3. Imaging of unfunctionalised Ag nanoparticles with varying scan parameters.** Top two rows: The top left image was recorded with a gain of 5%. For each consecutive image (i.e. moving along the rows from left to right), the gain was incremented by 1%. Each image is 8 nm wide, and all were recorded with a tip speed of 38 nm/s. Bottom two rows: Trace (third row) and retrace (bottom row) image of Ag nanoparticles, upwards scan direction. At the point marked by an arrow in both images, the scan speed was reduced from 514 nm/s to 195 nm/s, causing a significant reduction in stripe width (indicated with red arrows; these arrows also indicate scan direction). Soon after, the gain was reduced from 22% to 10% and the stripes disappear (gradually decreased in the lines marked by the green double-headed arrow). Both images have a width of 50 nm.

To put the analysis of the feedback noise contributions on a much sounder quantitative footing, we have performed Fourier transforms of the fast scan lines of the tunnel current images associated with Figure 3 of Jackson *et al.* 2006 [24], as feedback noise should dominate in the current channel. Feedback noise will also be aligned along the fast scan direction. We then combined the power spectra from each of the scan lines to locate the peak spatial frequency and the full-width at half maximum (FWHM) of the peak in the Fourier spectrum. The FWHM of the spectral peak gives a good measure of the range of frequencies which can arise from feedback noise. Plotting these spatial frequencies along with digitised data from Figure 3 Jackson *et al.* 2006 [24], as shown in Figure 4, it is possible to show that *all* of the quoted ripple spacings fall within the broad background noise measured for the whole image, and are hence not significant. One should also note in Figure 4 the systematic overestimation of the noise spatial frequency and underestimation of the noise error bars in the analysis by Jackson *et al.* 2006 [24], further demonstrating the inaccuracy of measuring ripple spacings by counting relatively few pixels.

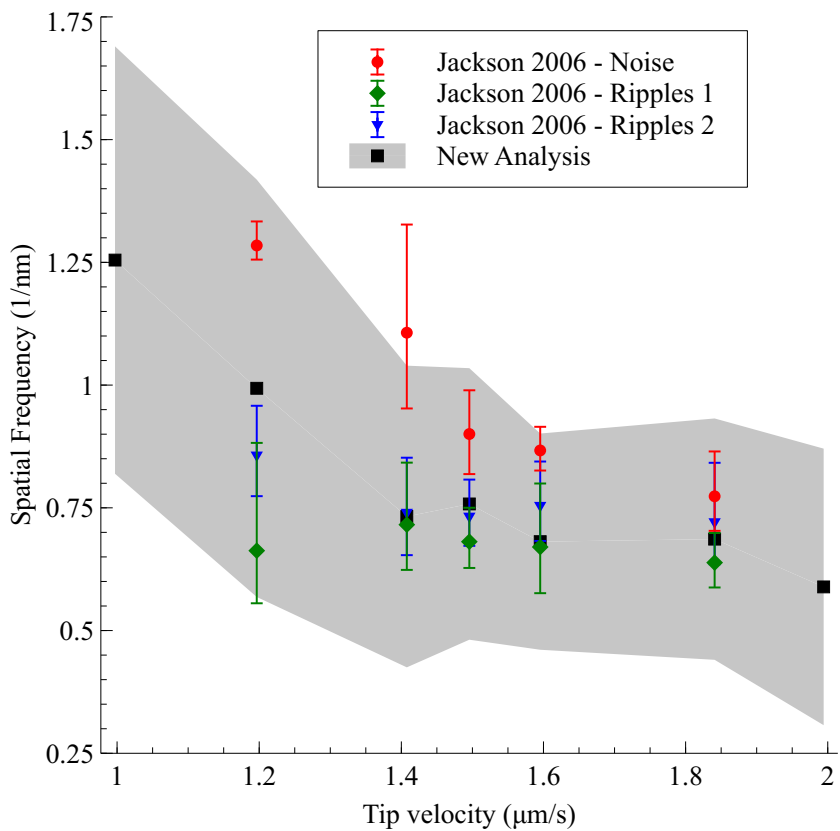

**Figure 4. Reanalysis of the data for Figure 3 of Jackson** *et al.* **2006 [24].** The black squares represent the peak frequency in the Fourier spectrum of the tunnel current images, while the grey area represents the full width at half-maximum (FWHM) of the peak in Fourier space. Red circles are digitised data from the noise spacings presented in Figure 3b of Jackson *et al.* 2006 [24]. Green diamonds and blue triangles are digitised data from the ripple spacings presented in Figure 3(b) of Jackson *et al.* 2006. *All ripple spacings fall inside the spatial frequency band of the error signal.* The first and last point represent images archived by Stellacci *et al.* along with the data for Figure 3 of Jackson *et al.* 2006 [24], but which were not analysed in Jackson *et al.* 2006. The full method and code used to generate this figure are given in the Supplementary Information.

Jackson *et al.* 2006 [24] also state that the gold foil substrates used in the work have "*curvature comparable to that of the nanoparticle core*". This begs the question as to just how some areas were objectively defined as the surface, and thus exhibited feedback noise, while others were defined as nanoparticles with molecular resolution. Furthermore, the areas defined as nanoparticles in the images do not show clear striped domains. Instead, they show a disordered noisy pattern.

For all of these reasons, the conclusions drawn by Jackson *et al.* 2006 [24] regarding their ability to distinguish true topographic "stripes" from feedback loop artefacts are entirely unreliable. Before we move away from the discussion of Jackson *et al.* 2006 [24], we would like to bring to the reader's attention more evidence of over-processed images. Figure 4 of that paper has been manipulated such that the contrast has only very few levels and the image appears as more of a contour map than a real STM image. Equally striking is Figure 9b of Jackson *et al.* 2006 [24]. In the context of discussing the orientation of stripes while rotating the scan angle, the inset, which is referred to as "Enlarged image of the same nanoparticle as in (a)", is actually an angled 3D rendering of the image, thus distorting the scan angle and providing an unfair comparison. Figure 8d of Jackson *et al.* 2006 [24] has lines drawn to "guide the reader's eye" to the direction of the stripes, arguing they are not aligned to the scan direction. This, however, masks the contrast and is yet again very misleading. An examination of the region which was enlarged simply does not show clear stripes in this direction.

We now turn to the second paper from Stellacci and co-workers which "critically assessed" the STM evidence for the striped morphologies: Hu *et al.* [36]. This paper solely concentrated on statistical analyses of their STM data. In common with Jackson *et al.* 2006 [24], the central claim is the ability to differentiate between stripes formed from feedback noise and those arising from real topographic features. This was based on a "rigorous" statistical analysis, where ripple spacings — again measured by eye, and thus subject to the same observer bias present for the analysis in Jackson *et al.* 2006 (Figure 3) — were compared to noise spacings while the tip speed was changed.

In one aspect the methodology is improved from that in Jackson *et al.* 2006 [24], in that separate images were used for topographical ripples and noise. The experimental methodology nonetheless still suffers from various other fundamental flaws. For a rigorous comparison, as the authors claim, each image taken at varying tip speeds should be of the same sample area, with the same scan size, and with the same feedback gain settings. The gain settings are especially important as we have shown above that the ripple spacing depends on feedback gains as well as tip speed. The archived data provided for the Hu *et al.* [36] paper has a selection of non-consecutive images, with sizes ranging from 2 to 300 nm, each with different gains, of different areas of the sample, or often of entirely different samples. As so many experimental variables are changing it is impossible to isolate the effect of tip speed, especially as gains have a pronounced effect on stripe width (Figures 2 and 3).

We also take issue with misleading descriptions of data acquisition in Hu *et al.* [36]. When describing the influence of tip speed on ripple spacings it is stated that "Many images are analyzed at varying tip speeds. In some cases we have analyzed as many as 10 images". Originally we understood this to mean that each speed had as many as 10 images, and the resulting data point was an average. After receiving the archived data (along with private communications with the research group [44]) we have found that each data point (i.e. for a given tip speed) is instead from a single image. The "10 images" refers simply to ten separate data points, each with different speeds, taken on different areas of the same sample (with other changing experimental conditions). Furthermore, the number of data points, indicated for different samples, does not agree with the number of images provided: at times the archive is missing images, and for other samples, more images are provided than were measured.

## Pixelation, offline zooms, and interpolation

Cesbron *et al.* [35] identified that the striped features observed for mixed-ligand-terminated particles, as of 2012, were all aligned with the scan direction. This was used as a central argument of the paper to suggest that the stripes were not true features but artefacts from feedback loop ringing (The analysis of the raw data described above confirms this interpretation). In response to Cesbron *et al.*'s criticism, Yu and Stellacci [47] provided examples of stripes which were not aligned with the scan direction. Those particular images, however, while not exhibiting feedback loop instabilities, suffer from a combination of poor experimental design, flawed analysis techniques, and strong observer bias, which we also critique in depth in the following.

The images in Figures 3 and 4 of Yu and Stellacci [47] were recorded using an Omicron micro-STM under UHV conditions, a microscope capable of acquiring high resolution images of just a few nm across, and of providing atomic resolution on flat surfaces [48]. The images, however, were acquired using a scan area of $80 \times 80$ nm$^2$ ($400 \times 400$ pixels), on nanoparticles with a diameter of order 4–6 nm. No data were presented where the scan range was decreased to record high-resolution images. Instead, zooms were yet again performed offline. Yu and Stellacci presented further enlarged figures showing single nanoparticles which were of order 30 pixels across, with a particle itself having a diameter of order 20–30 pixels. These images were then (inadvertently) interpolated via an image analysis package to show smooth "stripe" features. The "stripes", however, arise from as few as 2–3 noisy pixels in the original, uninterpolated, image. As such, this is a fascinating example of how improper image acquisition and analysis, coupled with observer bias, can lead to the observation of features which do not exist.

The human brain is well known to recognise expected patterns were none are present [49, 50]. A particularly important example is the observation of *perceived* correlated features in Poisson point distributions (where no spatial correlation exists). To ascertain whether stripes are present, therefore, it is important to carry out a rigorous quantitative analysis. Although, to the very best of our knowledge, no high resolution images were ever taken by Yu and Stellacci, many low resolution images of the same sample area were acquired (which the corresponding author kindly sent to us for analysis). These repeated images of the same sample area can be used to demonstrate that the stripes, which are claimed to be present in Figure 3 and 4 of Yu and Stellacci [47], arise from a misinterpretation of random noise.

First, we note that the 'full' images in Figure 3 of Yu and Stellacci are digital zooms ($\sim 40 \times 40$ nm$^2$) of the original $80 \times 80$ nm$^2$ images. A cursory analysis shows that the original images shift only by 4–5 nm between scans. Thus, it would have been easy for the authors to locate precisely the same particles and show that, if the features did indeed arise from organisation in the particle ligand shell, the stripes for all of the particles remained unchanged as the scan speed varied. This is not what is included in the paper (for reasons which will become clear). Instead, for each scan included in Figure 3 of Yu and Stellaci [47], the selected nanoparticles are different. This 'cherry picking' of the 'best' particles is used to suggest consistency between the images when none is present. To highlight this, we show in Figure 5 the summation of a $100 \times 100$ pixel section of all five images from both Figures 3 and 4 of Yu and Stellacci (trace and retrace, in total a sum of ten images), where these images have been aligned using cross-correlation. If the stripes identified by Yu and Stellacci [47] arise from a source other than noise they should still be visible in the sum of the images (The summation of data in this manner is a basic protocol in experimental science to increase signal-to-noise ratio). The summed data, however, shows smooth particles and the inescapable conclusion is that the stripe features arise solely from noise.

Yu and Stellacci used the same set of images to suggest that identical features can be recognised after a scan rotation. First, if features are supposedly visible in consecutive images after a rotation, it cannot simultaneously be argued that the ligands (or particles) shift sufficiently from scan to scan such that the stripes cannot be resolved in consecutive images. Let us assume, however, that we adopt the argument, entirely lacking in self-consistency, that features on the same particle which rotate as a function of scan rotation somehow are not present from scan to scan. Those features should nonetheless be present in the retrace image, which is taken *at the same time as the trace image.*

Figure 6a), (c), and (f) show images from Yu and Stellacci with arguably the strongest contrast of all of the features presented in that paper. Figure 6b) is a $205 \times 205$ pixel section of the raw data. In order to recreate the contrast in (a) we have flattened with a second order polynomial and then over-saturated the image by running the colour range from 35% to 75% of the full data range, before finally interpolating up to 820 pixels. Figure 6d) shows a crop of Figure 6b) showing approximately the same area as in (c), whereas (g) is the raw uninterpolated image where the individual pixels may be discerned. (The colour range is again reduced to increase contrast). Figure 6e) and (h) are equivalent to Figure 6d) and (g) respectively taken for the simultaneous retrace where we note that the stripes are not present on this

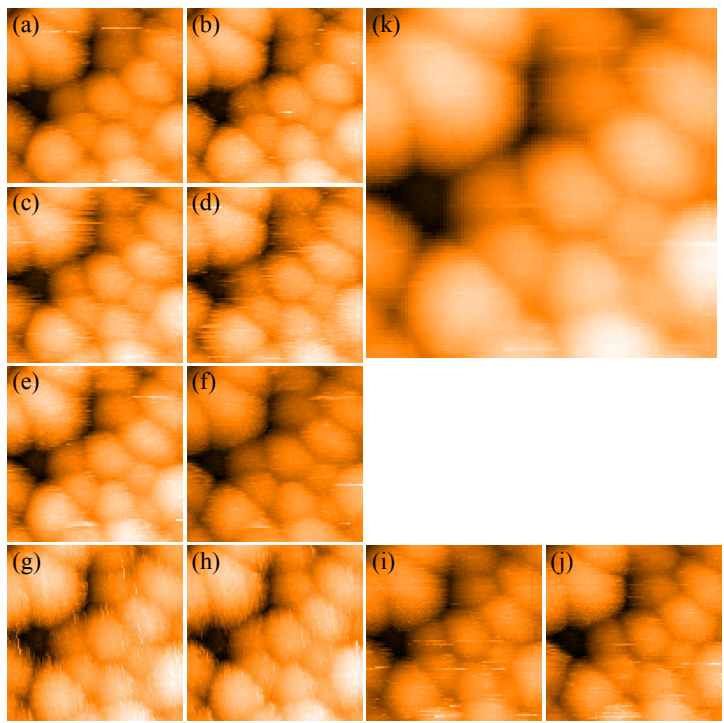

**Figure 5. Arithmetic addition of images from Yu and Stellacci [47].(a)-(j)** Images of the same set of nanoparticles taken from each of the five trace and five retrace images provided by Yu and Stellacci. (a,c,e,g,i) are the trace images, while (b,d,f,h,j), respectively, are the corresponding retrace images. **(k)** Arithmetic addition of all 10 images. Note that the particles in the summed image appear entirely smooth, indicating that the features designated as stripes by Yu and Stellacci arise from noise and not real topographic structure on the nanoparticles. All images are 20 nm wide.

image. We again must conclude that the "stripes" identified by Yu and Stellacci arise purely from a combination of noise and strong observer bias. In the Supplementary Information a program is included which allows the user to browse the trace and retrace images from Yu and Stellacci [47] (both raw and interpolated) simultaneously to show that this result is consistent across all particles and all images.

**The state of the art in resolving "stripes" — Data published in 2013**

Three further papers claiming to have found evidence for stripes in STM images have been published in 2013. We start with a consideration of Ong *et al.* [38]. This work details new data acquired by three separate STM groups (including that of Stellacci) from the same samples [51]. The images collected are certainly of significantly higher resolution and of higher quality than images presented in earlier work. Despite this increased resolution, however, there is a pronounced absence of stripes in the images presented by Ong *et al.* [38].

It is particularly instructive to compare the high contrast stripes presented in Figure 1a) with the STM images of mixed-ligand nanoparticles acquired by Ong *et al.*, which are shown in Figure 7a) and (b). These latter images reputedly show individual ligand head groups arranged in stripe-like domains. For further comparison, Figure 7c) shows an image of a homoligand nanoparticle from the same paper; stripes are supposed to be absent from homoligand particles.

Ong *et al.* [38] use the persistence of features in trace and retrace images, and in consecutive images, as evidence that the features in the images are real. It is worth noting that we used precisely this approach in the preceding section to show that the stripes in the STM images of Yu and Stellacci [47], published less than a year before Ong *et al.*'s work, are clearly artefactual. However, the persistence of features from scan to scan in the data shown in Ong *et al.* is somewhat irrelevant: the scanning protocol provides no support for the presence of a striped morphology in the shell of mixed-ligand terminated particles, because the evidence for the presence of stripes in the STM data is far from compelling. Nonetheless, the data of Ong *et al.* [38] highlight an important misconception in the analysis of SPM images which we feel needs addressing before we critique that paper in detail.

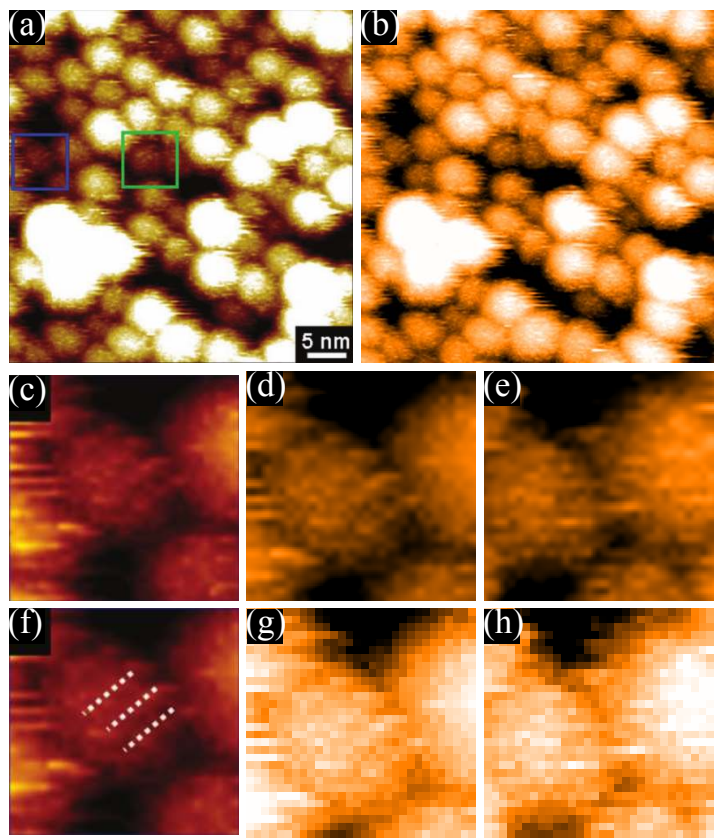

**Figure 6. Reanalysis of data from Yu and Stellacci [47].** **(a)** Image as presented in Yu and Stellacci; **(b)** A $205 \times 205$ pixel section of the raw data which has been processed with second order background subtraction, the colour range reduced to just 40% of the original range, and the number of pixels interpolated to best match the image shown in (a); **(c)** Enlargement of region highlighted by a blue square in (a); **(d)** Zoom of a section of the image shown in (b) taken after interpolation and colour saturation; **(e)** Retrace image acquired simultaneously with (d); **(f)** Image shown in (c) but with the stripes identified by Yu and Stellacci highlighted using dashed lines; **(g)** Uninterpolated zoom of the raw data showing the true pixelation. **(h)** Retrace image acquired simultaneously with (g). The "stripes" in (f) not only arise from a very small number of fortuitously aligned pixels, but they are not present in the retrace images shown in (e) and (h).

The difference between trace, retrace, and subsequent images is useful to identify feedback artefacts and noise-induced features. However, this approach simply cannot identify artefacts produced from tip-sample convolution. If the tip has a similar radius of curvature to features on the surface then convolution can be very pronounced [52]. This can even be used to produce images of a tip instead of the sample [32, 53–55]. For this reason, the 'internal' contrast of nanoparticles must be considered in the context of the apparent structure of neighbouring particles (or other surface features). Note that Figure 7b), for example (and unlike Figure 7c), shows an isolated particle with no surrounding nanoparticles with which to compare the internal structure.

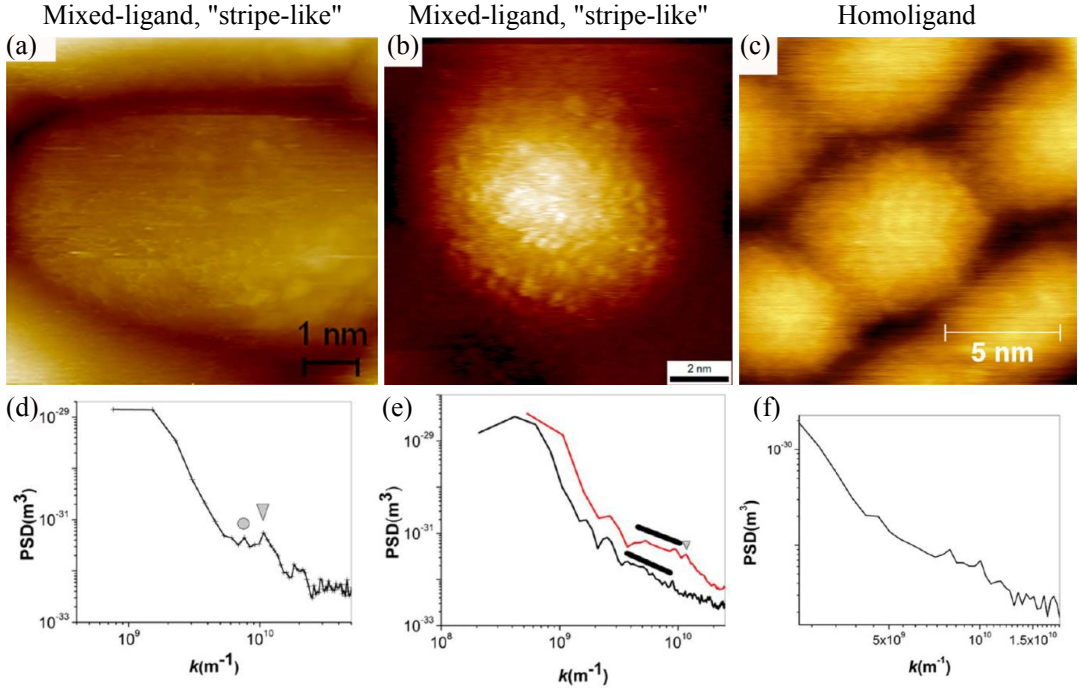

**Figure 7. Representative data from Ong** *et al.* **[38]. (a)** High-resolution STM image of an Au nanoparticle with a coating of 11-mercapto-1-undecanol and 4-mercapto-1-butanol, taken in UHV conditions at 77K. **(b)** High-resolution STM image of an Au nanoparticle with a coating of OT:MPA used in the original striped morphology paper (Jackson *et al.* 2004 [23]). **(c)** High-resolution STM image of homoligand nanoparticle with an OT coating. (a) and (b) allegedly show stripe-like domains while (c) does not. **(d–f)** Radially averaged PSDs from STM images of the same type of particles shown in (a)–(c) respectively.

To highlight the influence of the tip state on the apparent structure of nanoparticles, Figure 8 shows a series of images of the Ag nanoparticle sample, which was used for the loop-gain dependent studies shown in Figure 3. Each particle clearly exhibits detailed internal structure *which is entirely artefactual* and which, although being of the same general form across the image, varies somewhat in detail from particle to particle due to changes in nanoparticle structure, and thus the nature of the tip-sample convolution. In the row of images at the bottom of the figure we show how the apparent topography of just one of the nanoparticles varies as a function of the tip structure. There are a number of tip change events (red arrow) throughout the sequence shown in Figure 8, but it is important to note that during the intervals between the tip changes the images are entirely stable and checks of image "integrity" such as rotating the scan angle would show that the particle sub-structure behaved as one would expect real structure to behave. It is also interesting to note from Figure 8 that "Janus" nanoparticle [56] artefacts are very commonly produced in STM images due to tip structure (see, for example, the lower half of Figure 8a) and both Figure 8c) and (d)).

Returning to the discussion of Ong *et al.* [38], two methods were used to ostensibly distinguish striped morphologies. First, after plane-fitting the data, convolution with a 2D Mexican-hat wavelet (effectively a highly localised bandpass filter [57]) was used to highlight features of a specific chosen size [58, 59]. These were interpreted as ligand head groups. It is perhaps worth noting that the wavelet convolution used is described as a continuous wavelet transform. This is incorrect, as the frequency is not allowed to vary [59, 60]. Instead, a particular spatial frequency of the wavelet was chosen by the user. The highlighted features were located using watershed analysis, marked in the manuscript images, and shown to form clusters.

We make two key points regarding this analysis. First, using watershed analysis on structure highlighted with the type of convolution approach employed by the authors will locate features in almost any image if the settings are adjusted appropriately. More importantly, clustering of point-like features is expected for a random (Poisson) distribution [61, 62]. No attempts to analyse the spatial distribution of the features — via, for example, correlation functions or Minkowski functionals [63] — to assess the degree of randomness is made. As mentioned previously, careful quantitative analysis is essential as humans instinctively recognise patterns where no true spatial correlation exists [49, 50].

To highlight this problem, in Figure 9 we compare the distribution of assigned head groups and striped domains from an image in Ong *et al.* [38] with randomly positioned particles. Note how the eye can very easily be tricked into finding patterns in particles which have zero spatial correlation. The code used to generate the randomly distributed particles and the distribution from Ong *et al.* is given in the Supplementary Information.

The second method used in Ong *et al.* [38] to detect striped morphologies is to use a radially averaged 2D power spectral density (PSD) plot. The 2D PSD is the modulus squared of the 2D Fourier transform. A radially averaged PSD indicates the presence of oscillating features in *any* spatial direction. As this paper concentrates on images of single nanoparticles, where oscillations from stripes will have a particular orientation, radially averaging simply removes any directional information present in the 2D PSD. Figures 7 (d–f) correspond to radial PSDs of the same type of nanoparticle samples imaged in (a–c) of that figure respectively. (Note, however, that the PSDs are not taken from the images shown in (a–c)). The triangle and circle in Figure 7d) mark small peaks in the radial PSD when plotted on a logarithmic scale. These peaks are interpreted as corresponding to the spacing between head groups within stripes and the distance between stripes with distances of 0.59 and 0.83 nm, respectively. We note that even for a *square* grid of features, one would expect two peaks in a radial PSD corresponding to row spacings and diagonal spacings, with a ratio of $\sqrt{2} = 1.414$. The ratio between spacings in Figure 7d) is $0.83/0.59 = 1.407$ which agree with a square grid to 3 significant figures. We do not use this observation to imply that the features in the image are distributed on a square grid, but simply to point out that there are multiple possible interpretations of a radial PSD of point-like features.

The PSD analysis also suffers from other flaws. Ong *et al.*, use the line in Figure 7e) to define a wide peak corresponding to the distance between stripe-like domains. Remarkably, however, the two peaks present in Figure 7f), are not marked, despite being significantly stronger than those in Figure 7e). Those features are nonetheless mentioned in the text of the paper, where they are assigned to distances present in the randomly ordered ligand arrangement. This assignment begs the question as to why the peaks in Figure 7d) and e) could not arise from random ordering; why the full 2D data was not analysed to get directional information on these peaks; and why no mathematical analysis was applied to test for randomness in the located head group positions.

The radial PSD approach employed by Ong *et al.* [38], therefore, cannot be used to objectively determine whether stripes are present in the nanoparticle ligand shell. We now turn to a critique of the 1D PSD method used in a paper published shortly after that of Ong *et al.* where Biscarini *et al.* [39] apply a modified PSD method to quantitatively analyse both new and old STM images from Stellacci *et al.*. In Biscarini *et al.*'s case, a 1D PSD is acquired by calculating the PSD for each scan line in the image and averaging down the slow scan direction. This method will capture stripes aligned with the scan direction while stripes of spatial frequency $f$ misaligned by an angle $\theta$ will appear at a frequency of $f \cos \theta$. Thus, if 1D PSD analysis of this type is applied to an image with randomly aligned striped particles, one expects a broadened peak near the stripe spatial frequency (assuming that there is a sufficiently high number of particles in the image to produce a well-resolved peak).

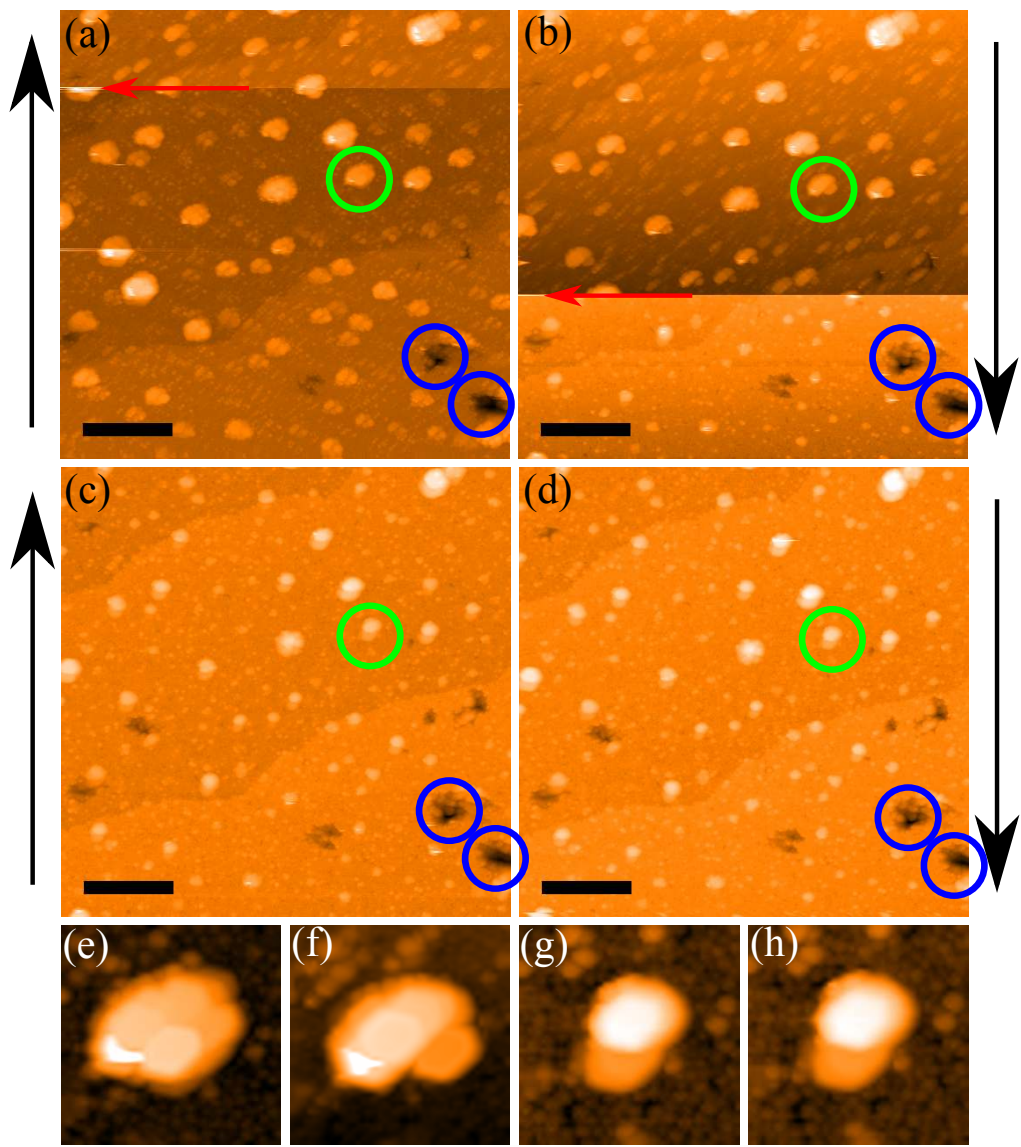

**Figure 8. The persistence of tip induced features on bare Ag nanoparticles.** Four successive images (a–d) with black arrows showing the direction of the slow scan. Tip change events, marked by a red arrow, change the apparent sub-particle structure of the bare nanoparticles. Note the persistence of the artefacts throughout the images. The tip state shown in (d) was persistent over many consecutive scans. The green circle identifies the same particle in subsequent images and (e–h) show offline (and interpolated) zooms of this particle from each of the images (a–d). Blue circles mark the same features in all images as a reference point to show the scan area is consistent. All scale bars in (a–d) are 30 nm. Minor contrast adjustment has been applied to images (a,b,d,e,f).

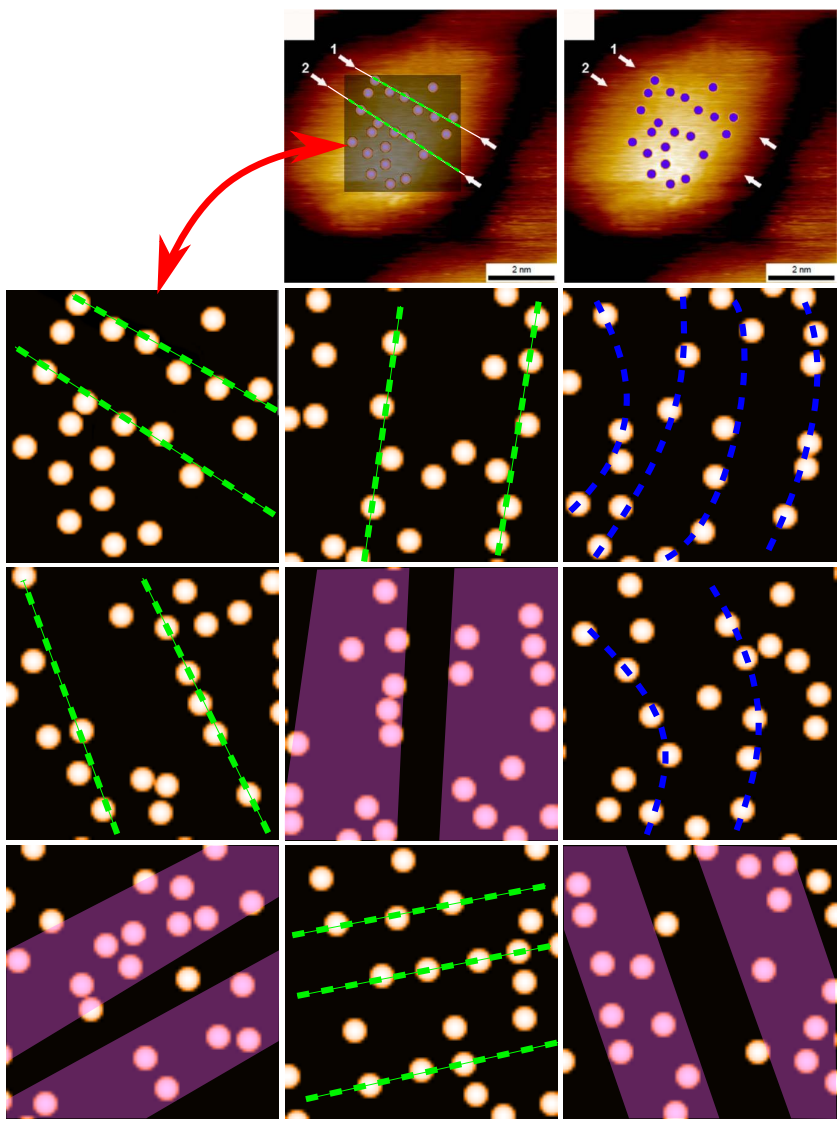

**Figure 9. Digitised position of ligand head groups and stripes identified by Ong et al. [37] as compared to eight sets of randomly distributed 'head groups'.** The top row shows the image in question from Ong et al. (upper right corner) along with a version of that image where we have superimposed a semi-transparent square and highlighted the 'stripes' identified by Ong et al. using green dashed lines. The original blue circles (right) are visible through the digitised head groups. The positions of the head-group features within that square, and the corresponding dashed lines highlighting the 'stripes', are then reproduced on a featureless background, as indicated by the red double-headed arrow. The other eight images in the figure for comparison show randomly distributed features. By either assigning straight lines (green), curved lines (blue), or stripe-like domains (purple) it is possible to guide the reader's eye to clustering in random features.

When plotting the 1D PSD on a logarithmic scale, Biscarini *et al.* [39] observe an initial plateau and shoulder arising from the characteristic size of the nanoparticles, followed by a decay, then a second plateau and shoulder, followed by another decay. The second plateau and shoulder is, rather precipitously, taken as evidence for the striped morphology. Little time is spent by Biscarini *et al.* [39] to determine that this shape cannot arise from other image features. We show in the following that the plateau and shoulder do not arise from stripes, but from a random arrangement of features on the nanoparticle surfaces.

Figure 10c) shows a simulated nanoparticle substrate. If stripes are present on the particles (Figure 10d–f)), then the expected broad peak forms in the 1D PSD Figure 10a). We also note that the stripes are clearly visible to the eye before the 1D PSD peak becomes noticeable. If, however, randomly positioned speckles (Figure 10g)) are added to the substrate (Figure 10h–j)), the plateau and shoulder observed by Biscarini *et al.* in the experimental data are produced. Indeed, Biscarini *et al.* observe a very similar plateau and shoulder for homoligand nanoparticles, but they argue that because the shoulder appears at a different spatial frequency this distinguishes it from the structure in the PSD arising from the stripe-like morphology. This is an entirely unwarranted conclusion to draw and begs yet another question: why does the presence of the plateau-and-shoulder structure in the PSD at a different spatial frequency not lead to the natural conclusion that the PSD points to the presence of a similar (random) morphology, but at a different characteristic length scale? Biscarini *et al.* [39] do not address this exceptionally important point.

In order to bolster their case that the STM images used for their analysis are artefact free, Biscarini *et al.* [39] fit the PSD to extract characteristic frequencies which should be unchanged under varying scan speed, similar to the analysis in Hu *et al.* [36], except using Fourier analysis. This analysis however is once again multiply flawed. First, where stripes are not clear to the eye (and because, as shown above, the 1D PSD cannot distinguish between stripes and other morphologies), even if the spatial frequencies are real, this does not represent evidence for a striped morphology. In addition, as for the data previously analysed in Hu *et al.* [36], and discussed above, due to the variation of multiple scan settings in addition to the scan speed the test is not rigorous.

An additional fundamental difficulty with the analysis presented in Biscarini *et al.* [39] is that the fitting procedure used to extract spatial frequencies from the PSD data is very far from robust. Furthermore, the description of the fitting process given by Biscarini *et al.* in their paper is misleading at times. We describe the difficulties with the fitting process in detail in the Supplementary Information. Here, we simply state the following: (i) there are seven free parameters in the fit. Multi-parameter fitting of this type is not at all well-suited to extracting reliable (and unique) spatial frequency values [64, 65], particularly when the fitting was carried out by Biscarini *et al.* [39] in the manner described in the Supplementary Information; (ii) sections of the PSD data were excluded from the fit by Biscarini *et al.*, without this exclusion being explicitly mentioned in the text of the paper [66]. Even if this would not be the case, the initial choice of fitting parameters can substantially bias the output of the fitting algorithm; and (iii) we have repeated the fits in MATLAB and find that in all cases warnings for poor convergence were given.

As a final note on Biscarini *et al.*, the PSD analysis is repeatedly argued as the best method for measuring image features as it contains the "whole information content present in the image", and as such, is unbiased. This shows a fundamental misunderstanding of Fourier analysis in that much of the information content of an image is contained in the phase components, and by taking PSD from the Fourier transform all phase information is lost. In addition, by *choosing* to average over a *particular* direction further information content in other directions is lost.

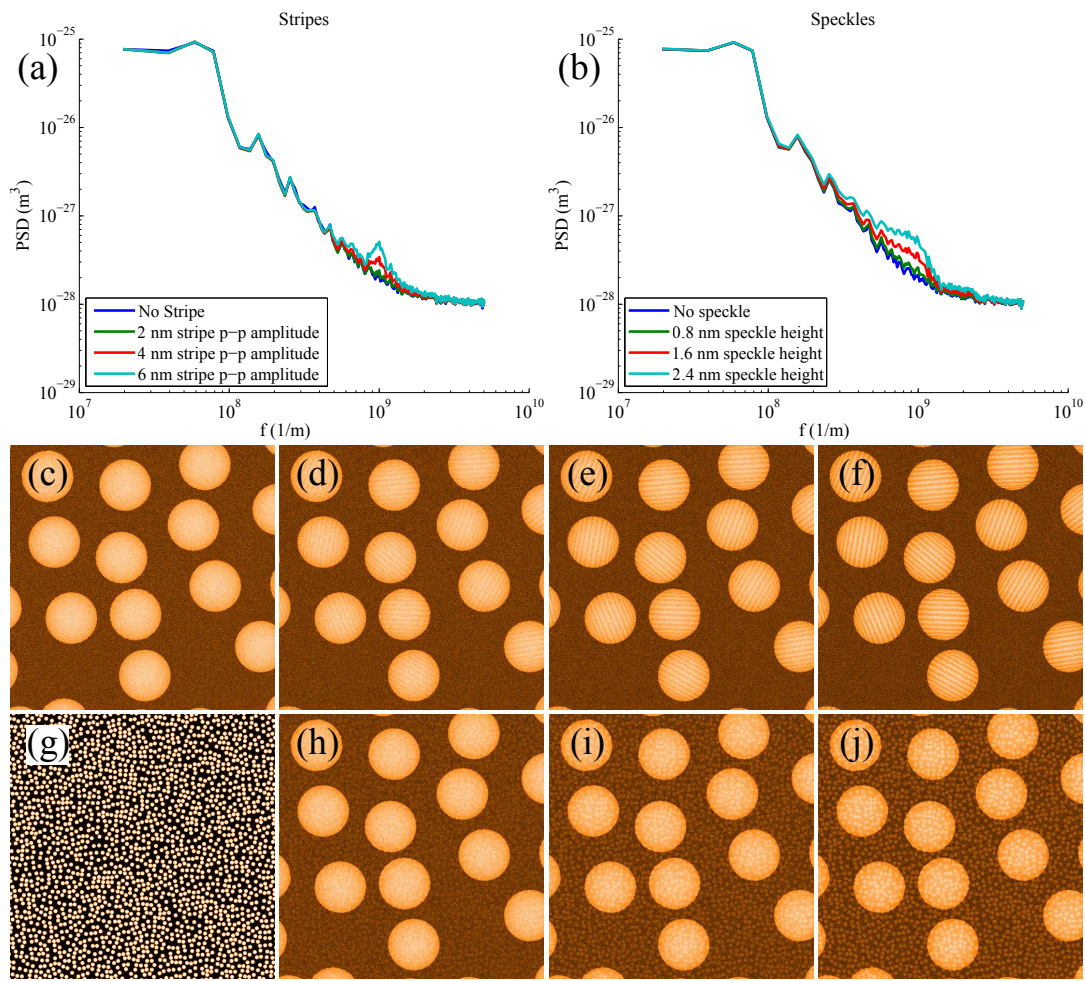

**Figure 10. 1D PSDs of simulated nanoparticle substrates. (a)** 1D PSD for nanoparticles for simulated stripes of increasing amplitude (see simulated images shown in c–f); **(b)** Equivalent to (a) but in this case for simulated nanoparticles covered in randomly positioned speckles (ligand head-groups (see images (h–j)). *The speckled images simulating a random distribution of head-groups yield the plateau and shoulder observed by Biscarini et al. [39] which were inadvertently assumed to represent the signature of a striped morphology.*; **(c)** Simulated flat surface with 10nm diameter spherical nanoparticles. **(d–f)** 1nm wide sinusoidal stripes are added to the surface of the nanoparticles (thus, they reduce in width at the edge) with peak-peak amplitudes of 2, 4 and 6 nm respectively; **(g)** Randomly distributed 'speckled' pattern of features 0.8 nm in diameter. **(h–j)** Images of simulated nanoparticles where the speckles in (g) have been added to (c) with heights of 0.8, 1.6 and 2.4 nm respectively. (c–f and h–j) have had identical white noise added for consistency and for a fair, unbiased comparison.

As the last paper to be considered in this section, we turn to Moglianetti *et al.* [40], where the role of scan rotation on liquid STM images of a new type of mixed-ligand-terminated nanoparticle (dodecanethiol: hexanethiol, 2:1) was studied. The PSDs of the STM images are also compared to data collected using small angle neutron scattering (SANS). The nanoparticles reportedly showed no striped morphology when imaged in ambient conditions while the liquid STM images presented instead are argued to show "clear stripe-like domains" for the particles.

Although Figure 1 of Moglianetti *et al.* [40] shows arguably the most convincing images of nanoparticle sub-structure we have seen to date in the work of Stellacci and co-authors (the persistence of features in the trace and retrace images is particularly compelling), the paper, far from demonstrating the presence of "clear stripe-like domains", provides no evidence for stripe formation. Once again, there is strong observer bias in the identification of "stripes". We suggest that the reader compare the dashed lines used to highlight the presence of "stripes" in Figure 1(e) of Moglianetti *et al.* [40] with those shown in Figure 9 above, where the head-group features are randomly distributed.

As the SANS data in Moglianetti *et al.* [40] are directly compared to the STM results and also involve an analysis of 1D PSD curves, we will discuss the SANS results in this section. The SANS data are, in essence, radially averaged scattering intensity plots which give a very similar information content to the radial PSD plots discussed above. The SANS data are interpreted as providing evidence for stripes in three ways. We shall take each of these points in turn. First, it is argued that the SANS data can be better fitted to a model for striped domains than for random particles. We refer the reader to Figure S5 to emphasise our point that fitting complicated features with a multi-parameter fit is highly open to interpretation and that the choice of fitting parameters can very easily skew the results. Second, 3D rendered nanoparticle surface morphologies were generated from the SANS data and are shown in Figure 5 of Moglianetti *et al.*. This figure shows patterns which, as discussed previously, are difficult to distinguish from random distributions. No attempt was made to quantitatively determine to what extent the morphologies differ from those which would statistically be expected from a random distribution of head-groups. Moreover, and importantly, these patterns differ rather dramatically from the ordered striped domains presented in the cartoons of a significant number of earlier papers from Stellacci and co-workers.

The third point regarding the SANS-STM comparison builds on the arguments re. PSDs outlined in relation to Figure 10 above. The SANS data were extrapolated to form a simulated STM image, and the 1D PSD analysis from Biscarini *et al.* [39] applied. As we have shown, the 1D PSDs calculated by Biscarini *et al.* [39] cannot distinguish stripes from other image features. Perhaps more importantly, the characteristic length extracted for the SANS-derived and STM PSDs differs by 50%. This discrepancy shows that either the PSD fitting is performing poorly, that the simulated image from the SANS has a different morphology or spacing from the true sample, or that the STM image has a different morphology or spacing from the true sample. It is perhaps most likely that there is a combination of all three effects.

Finally, it would be remiss of us to leave the discussion of Moglianetti *et al.* [40] without highlighting a troublesome misconception regarding STM image acquisition. In their paper, Moglianetti *et al.* [40] claim that *"as one rotates the image, the tip approaches the sample from different directions, this in turn leads to a change in image resolution, due to variation in the convolution conditions and the asymmetry in tip shape"*. This statement betrays a fundamental misunderstanding of STM operation. Artifacts from improper feedback settings will indeed depend on the scan rotation, but convolution effects result from the orientation of the tip relative to the sample. This does not change when the image is rotated via a change in scan angle: *neither the sample nor the tip is physically rotated.* Instead, the direction of raster scanning is changed. Any convolution effects from the tip are, therefore, expected to rotate with the image, as noted above in the context of the discussion of Figure 8.

## Assessment of evidence for nanoparticle stripes from techniques other than STM

In this section we will briefly critique the evidence for striped nanoparticles from techniques other than STM. These span nuclear magnetic resonance (NMR) spectroscopy, transmission electron microscopy (TEM), and computational simulations. The data from Fourier transform infra-red spectroscopy (FTIR) studies have not been considered, despite Yu and Stellacci [47] citing FTIR data in their response to Cesbron *et al.* [35]. This is because the paper cited by Yu and Stellacci explicitly states that FTIR can be used only to screen for phase separation, but cannot distinguish between striped and non-striped morphologies.

### Analysis of NMR spectroscopy data

Liu *et al.* [28] present a method using 1D and 2D NMR spectroscopy which they argue can identify the morphology of ligand shells for mixed-ligand nanoparticles. The core data centres around three types of MLN with binary ligand mixtures. All three contain diphenyl thiol (DPT) as one ligand. The first nanoparticle type has a diameter of 4–5nm, with 3,7-dimethyloctanethiol (DMOT) and DPT ligand mixtures that are assumed to form random ordering. A second type has a diameter of 2.2–3nm, with dodecanethiol (DDT) and DPT ligand mixtures that are assumed to form Janus nanoparticles. Finally, a type with a diameter of 4–5nm, also with a mixture of DDT and DPT, is assumed to have a varying patchy morphology, which exhibits stripes at 1:1 ratios.

For the development of the NMR methodology, the morphology of the MLNs is assumed to be already known from STM data. This is critical because, as we have discussed at length in the preceding sections, there is no evidence from the STM data to date that stripes form in the ligand shell. In addition, the STM images for the Janus nanoparticles clearly show pairs of separate nanoparticles which are close together, ringed as ovals and described as single nanoparticles. From the NMR data, no direct evidence for the existence of the stripes is presented. The question of the validity of the reasoning, however, is still relevant to the argument for or against the striped morphologies.

Unfortunately, we found the data yet again to be inconclusive, combined with some major flaws in some specific areas of analysis. For brevity we will only discuss the 1D spectra below, as this forms the core of the presented evidence. The 2D data are, however, discussed in detail in the Supplementary Information.

The primary information used from the 1D NMR spectra is the chemical shift of the aryl peak maximum. There are various pieces of information that are not considered or interpreted. In particular, the line caused by the alkyl ligands is not analysed, despite its changing position and pattern. In addition, linewidths and lineshapes are not analysed in any way (neither in the 1D nor in the 2D data), with the exception of a narrow aryl line. This line is interpreted under the assumption that the morphology is known to be striped, and via an indirect argument based on the reactivity of ligands in nanoparticles. Further details regarding this narrow aryl line are presented in the Supplementary Information files.

The model used to explain changes in the chemical shift of DPT assumes a linear change from the bulk chemical shift to the chemical shift of DPT surrounded by the other ligand as the ratio of the second ligand to DPT is increased. This relation is referred to as "trivial" with no consideration that the chemical shift can depend strongly on possible changes in the local ordering of the phenyl rings relative to each other or on the mobility of the thiols, which will change with varying ligand ratios. This is because ring currents in the aromatic rings of DPT cause a highly orientation-dependent shift of the $^1$H NMR resonances as a function of the proton position with respect to the ring [67]. Further problems exist with this model [68], which are again addressed in detail the supplementary information.

Assuming the validity of this linear model, Liu *et al.* [28] continue to derive an equation for Janus particles, which they refer to as "rigorous". However, at neither concentration limit does the equation tend to the expected values; this point is never addressed. The model is fitted to the experimental results, but close inspection shows that both initial and final point are below the fit, with central points above. This trend in the residuals strongly suggests that the model does not fully explain the data. Upon reading the full text it becomes clear that to generate this fit the second point was arbitrarily designated as an outlier to increase the $R^2$ value. An $R^2$ of 0.976 is used to suggest the model "provides excellent agreement" with no mention of the clear trend in the residuals [64]. In the supplementary information, we derive a revised model for Janus particles which provided a more accurate fit without any data exclusion. Our model still falls short of a rigorous model as it fails to converge if the mole fraction of the ligands being detected falls below the value necessary to maintain two bulk regions. Then all of the corresponding ligand molecules are located in the interface region, for which case the model is not designed to make any predictions. The point is raised not to dispute the evidence for the presence of Janus particles (although the STM data are far from compelling), but simply to demonstrate further evidence of careless data analysis.

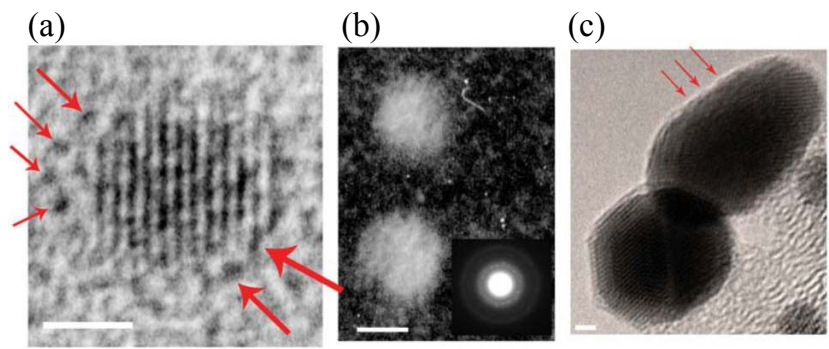

**Figure 11. TEM data of OT:MPA-coated nanoparticles from Figure S2 of Jackson** *et al.* **2004 [23]. (a)** Red arrows indicate dark features surrounding the nanoparticle which have been interpreted as MPA head groups. Such features commonly arise from TEM defocus, and even if real are not arranged in striped domains. **(b)** Dark-field TEM image with inset power spectrum. **(c)** Further TEM image of mixed-ligand nanoparticles with red arrows supposedly indicating sinusoidal features. Neither (b) nor (c) show any evidence for an ordered striped morphology.

The key conclusion of our re-analysis of the NMR data, however, is that the evidence presented for striped morphologies is exceptionally weak. Liu *et al.* [28] suggest that for patchy nanoparticles with stripes around 1:1 ratios, the chemical shift should vary as a sigmoidal function for increasing concentrations of DPT. This reasoning is not explained in their paper. As the change in chemical shift is dependent on the complex and unknown evolution of the patches a sigmoidal function cannot be assumed *a priori*. Similarly, no justification that other morphologies could not produce a sigmoidal function is given. In addition, the results do not unambiguously show a sigmoidal pattern. Instead, up to a DPT concentration of about 60% the chemical shift changes very little, followed by an almost linear reduction towards the bulk value. These data could also be equally well explained by the formation of small circular patches of DPT among DDT. As the concentration of DPT is increased more patches of similar size are generated, until a critical point is reached where the patches coalesce (see supplementary information for more details). Liu *et al.* [28] instead use the large uncertainties in the measurement to claim that the straight lines are not statistically significant and that the true dependence may well be sigmoidal, and hence the data would be in "excellent agreement" with the striped model. This argument can be used to claim that the data do not preclude the possibility of a striped morphology, yet cannot be used as direct evidence in favour.

### Analysis of TEM data

The TEM data cited as evidence of ligand stripes comprises just three images of OT:MPA MLNs in the supplementary information of Jackson *et al.* 2004 [23], Figure S2, reproduced in here in Figure 11. Taking (b–c) first, a dark-field and a bright-field image respectively, each shows two nanoparticles and neither show any evidence of stripes. The red arrows in (c) are reported to show sinusoidal features at the edge of the particle. For (a) single dark features around the particle are indicated with arrows, and were assumed by Jackson *et al.* [23] to be single MPA head groups. Our objection to this evidence is two-fold. First these features are of similar size to features in amorphous background, yet darker. A ring of such features is often seen in TEM images of bare nanoparticles [69–72], and can be enhanced or removed by varying the defocus [73, 74]. More fundamentally, even if these 6 to 7 features did represent MPA head-groups their ordering is in two groups, which may suggest phase separation, but no striped morphology is observed.

## Comment on computer simulations

To discuss the evidence for striped morphologies from computer models one must understand the role of simulations as an aid to understanding experimental data and making predictions from theories [75]. Simulations all come with their own advantages and difficulties, and, depending on what information is desired, different methods are applicable. For predicting structures, methods closest to *ab initio*, such as DFT, are usually preferable. Such simulations are computationally expensive and thus are only performed on relatively small numbers of atoms. Statistical methods such as Monte Carlo simulations [76] [77], or semi-classical approaches such as molecular dynamics [78], are less expensive and thus larger systems can be studied, but at the cost of decreased accuracy.

The bonding of thiols to Au surfaces is still not completely solved [79], but our understanding has improved vastly since the original simulations of striped morphologies on Au nanoparticles [29]. The prevailing view was that the thiols bond through the sulphur to the Au surface at a specific site [80]. More recent studies, however, indicate that thiols bond as Au-adatom-dithiolate structures (R-S-Au-S-R), with strong supporting evidence from DFT simulations [79], STM on Au surfaces [81], and on Au nanoparticles via x-ray diffraction [82]. Further DFT and XPS studies have shown variations in binding energy arising from interactions between dissimilar thiols [83].

The simulations presented as evidence for the striped morphology use a mesoscale simulation called dissipative particle dynamics [84]. Here intramolecular interactions are modelled as harmonic springs [29]. Intermolecular interactions are treated as harmonic potentials with the model parameters chosen to have a higher repulsion between atoms on unlike molecules. Ligand-Au bonding is not modelled. Instead, constrained dynamics are used to confine the head group to a sphere. This form of large scale simulation, due to the simplicity of the interaction and the unknown accuracy of the chosen parameters, cannot be used to reliably predict the complex structures on coated nanoparticles. It is instead used to search for experimentally known structures. Once these structures, and their evolution under changing conditions, can be matched to the outputs of the simulation it is possible to extract theoretical understanding of the observed structures. Further simulations were also performed using molecular dynamics with a similar constrained geometry, and selected potentials instead of repulsion parameters [29].

This approach to modelling not only simplifies bonding and molecular interactions, it also simplifies the structure of the nanoparticle itself. Nanoparticles capped in thiols are known to be more spherical than bare nanoparticles due to thiol interaction [85], but faceting is still present on the nanoparticles [82]. In addition, it is known that thiols modify gold surfaces [79] and nanoparticles [78] during the formation of self-assembled monolayers. Furthermore, the simulations only deal with the rearranging of randomly ordered thiols, not the posibility of structures arising from selective adsorption or ligand exchange [86,87].

These criticisms of the simulation are not meant to suggest that the simulation was poorly performed or is unjustified due to its simplicity. If a simple simulation can accurately describe and provide insight into experimentally observed behaviour then it is a valid simulation. However, *if the experimental evidence for the structure is called into question it is tautological to use a simplistic simulation designed to understand this structure as evidence that the structure itself does exist.*

# Conclusions

We have critiqued and re-analysed the extensive series of papers from Stellacci *et al.*, which argue that stripes form in the ligand shell of appropriately functionalised nanoparticles. The experimental evidence required to justify the claim of striped morphologies is lacking. Moreover, the majority of the published data suffers from rudimentary flaws due to instrumental artefacts, inappropriate data acquisition and analysis, and/or observer bias. The first paper claiming to resolve ligand stripes, Jackson *et al.* 2004 [23], shows features which arise from feedback instabilities and which can be reproduced on bare nanoparticles. Jackson *et al.*'s results were supplemented with papers which attempted to differentiate between artefacts and true nanoparticle topography on the basis of the variation of scan parameters. The methods used in these studies are far from rigorous, as multiple conditions changed between images. Moreover, the investigators hand-picked which features were to be analysed as artefacts and which were 'true' stripe features. Recent STM data, collected in collaboration with other SPM groups, despite being taken at significantly higher resolution shows a significant decrease in sub-nanoparticle contrast. The reduction in contrast is so strong that the stripes cannot easily be recognised in real space. To investigate the stripes Fourier space analysis has therefore been applied. We show, however, that the Fourier space techniques which have been employed are unable to reliably discriminate between stripes and other morphologies. Finally, the quantitative methods, which have previously been developed for extracting spatial frequencies from the resulting Fourier space data, are fundamentally flawed as they rely on a multi-parameter fit, which is highly sensitive to the initial, user-defined, fitting parameters. On the combined basis of our analysis of the flaws in the scanning probe studies and our criticisms of the evidence from other complementary techniques, we conclude that no reliable evidence has been presented to date for the presence of ligand stripes on mixed-ligand nanoparticles.

# Acknowledgments

The authors would like to thank F. Stellacci and F. Biscarini for sharing some of the raw data requested for re-analysis in this paper. We also thank David Fernig (Institute of Integrative Biology, University of Liverpool) and Richard Woolley (School of Physics and Astronomy, University of Nottingham) for many helpful discussions throughout the production of this manuscript.

# Critical assessment of the evidence for striped nanoparticles: Supplementary Information

Julian Stirling[1,*], Ioannis Lekkas[1], Adam Sweetman[1], Predrag Djuranovic[2], Quanmin Guo[3], Josef Granwehr[4], Raphaël Lévy[5], Philip Moriarty[1]

1 School of Physics and Astronomy, The University of Nottingham, University Park, Nottingham, NG7 2RD, United Kingdom

2 Department of Materials Science and Engineering, Massachusetts Institute of Technology, 77 Massachusetts Avenue, Cambridge, MA 02139, USA

3 School of Physics and Astronomy, University of Birmingham, Birmingham B15 2TT, United Kingdom

4 Sir Peter Mansfield Magnetic Resonance Centre, School of Physics and Astronomy, The University of Nottingham, University Park, Nottingham, NG7 2RD, United Kingdom

5 Institute of Integrative Biology, Biosciences Building, University of Liverpool, Crown Street, Liverpool, L69 7ZB, United Kingdom

∗ E-mail: rapha@liverpool.ac.uk

† E-mail: Philip.Moriarty@nottingham.ac.uk

# Contents

# 1 Overview of supplementary information

This supplementary information file details how to use the code (available from Reference [1])[1] to generate the figures presented in the main paper. It also contains extra information on the flaws in both the NMR spectroscopy anaylsis of Liu *et al.* [2] and Biscarini et al.'s [3] fitting of 1D power spectra curves which was not included in the main text for brevity.

# 2 Analysis of Figure 1 from Jackson *et al.* 2004 (Figure 1)

Figure 1 was created using raw data from Jackson *et al.* 2004 [4], part of the public archive released by Stellacci and co-workers in May 2013. This file is located at: `Public Data/Nature Materials 2004/Fig 1 original rippled files (ajs1)/raw files/npmono_gold_aj.006`

The full code to generate Figures 1(a,b,d,e) is provided in the file `NatureMat2004/Cut_n_Filter.m`. Figure 1(c) is from Jackson *et al.* 2004.

# 3 Feedback loop instabilities (Figure 2)

A real STM controller records the tunnel current at a specific sample rate rather than continuously, and all feedback calculations are discrete rather than analytical. As such, we argue it is most appropriate to model the STM system numerically.

For this we have decided, for speed and simplicity, to simply feed back on the height error, rather than convert a height into an exponentially decaying current and then take the logarithm. For this simulation the heights and gains are arbitrary. An algorithmic explanation of the feedback simulation is provided as pseudocode in Algorithm 1. The full simulation can be run from the file `Feedback/RunSimulation.m`, which requires the SPIW MATLAB toolbox [5], and the provided functions:

- `Feedback/GenNanoparticle.m`

- `Feedback/placeinpos.m`

- `Feedback/SimulateSPMFeedback.m`

Navigate to the Feedback directory before running. Run `doc functionname` for the help file of a particular function.

Final images are bicubically interpolated to a higher number of pixels to match the presentation of Jackson *et al.* 2004 [4]. In the folder we provide example outputs for those who do not have MATLAB. An example output is shown in Figure S1. For these outputs the top image shows the surface, the second row show the images before interpolation, and the bottom row shows the final interpolated images. Feedback settings are provided in the output image title.

An advantage of this numerical simulation is that it allows us to more accurately replicate the conditions of a real STM. Arbitrary topographies can be used, allowing us to build nanoparticle surfaces to scan, and add normally distributed noise to each measurement to simulate electrical noise. There are options to change the proportional and integral gains, the scan speed, and the set-point height. Additional options include changing the sample rate of the STM controller and changing the amplitude of the normally distributed noise. Finally one can turn on wind-up protection mode where the integral term in the feedback controller is reset after each pixel. (This mode tends to remove stripes or, in the case of very high integral gain, to 'lock' stripe widths to 1 pixel.)

---

[1] All code is released under FreeBSD licence, licence is included in root directory.

---

**Algorithm 1** Pseudocode for STM simulation algorithm. Braces indicate comments.

---

**Require:** The topography *topog* (an $N \times N$ array), time per line *Tline*, Sample Rate *SR*, Set-Point *SetPoint*, Proportional gain *Pgain* and Integral gain *Igain*, Wind-up protection (boolean) *Protect*.

**Ensure:** The simulated STM outputs *scan* and *scanR*.

1: $spp = \text{ROUND}(Tline^*SR/N)$ {Calculate the number of PID iterations taken per pixel}
2: $height = topog(1,1) + SetPoint$
3: I=0 {Initialise integral term}

4: {Using loops to simulate raster scan}
5: **for** $m = 1$ to $N$ **do**
6:     **for** $n = 1$ to $2^*N$ **do**
7:       **if** Protect **then**
8:         $I = 0$ {Reset integral term if wind-up protection is on.}
9:       **end if**
10:       {Calculate if on trace or retrace}
11:       **if** $n > N$ **then**
12:         $n2 = 2^*N + 1 - n$; {If on retrace, new fast scan position is calculated from $n$}
13:       **else**
14:         $n2 = n$; {If on trace fast scan position is simply $n$}
15:       **end if**
16:       **for** $i = 1$ to $spp$ **do**
17:         {Calculating error and adding normally distributed noise}
18:         $err = topog(m, n2) + SetPoint - height + \text{RANDN}$
19:         $I = I + err$
20:         {Using feedback to adjust height.}
21:         $height = height + Pgain^*err + Igain^*(I/SR)$
22:       **end for**
23:       {Scan data is the last height, must be written to either *scan* or *scanR* depending on if tip motion is trace or retrace}
24:       **if** $n > N$ **then**
25:         $scanR(m, n2) = height$
26:       **else**
27:         $scan(m, n2) = height$
28:       **end if**
29:     **end for**
30: **end for**

31: **return** *scan*, *scanR*

---

# 4   Parameter dependent imaging (Figure 3)

Figure 3 shows results for data collected on an Omicron LT at 77K in UHV on bare Ag nanoparticles as described in the main text. The raw data for all images in this Figure is provided in the folder `NewData/RawData`. The full experimental data collected is not provided due to its large size (over 5 GB) but is available on request.

The first two rows show consecutive retrace-down images with incremented gain between each image. This part of the figure is generated by the file `NewData/IncGain.m`. All images have been bi-cubically interpolated to four times the original pixel density in both dimensions. In addition, the particles have been aligned using cross-correlation to correct for drift between images. To compare the trace and retrace of a similar data set, see the file: `NewData/IncGain2.m` and the image output: `NewData/IncGain2.png`.

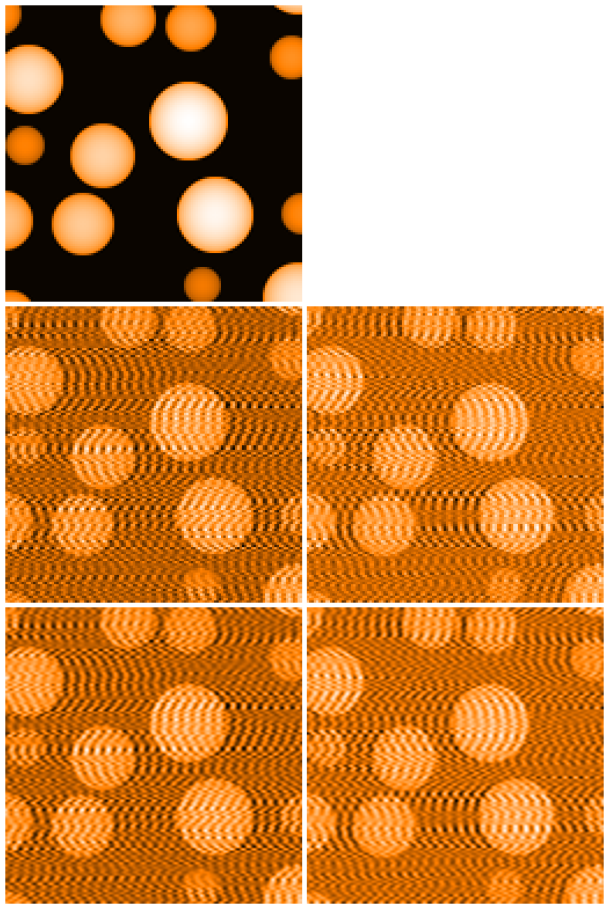

**Figure S1.** Example output from the program `Feedback/RunSimulation.m`. Top row shows the surface. Middle row trace (left) and retrace (right) before interpolation. Bottom row trace (left) and retrace (right) after interpolation.

# 5   Reanalysis of Jackson *et al.* 2006 (Figure 4)

Images from Jacskon 2006 were provided in the archived data provided by Stellacci and co-workers and can be found in the directory: `Public Data/Fig.3 Jacs 2006`. The analysis which produced Figure 4 is provided in three scripts which should be run successively:

1. `NewData/LineByLineAnalysis.m`

2. `NewData/make_compare_graph1.m`

3. `NewData/make_compare_graph2.m`

The code is well commented, but we provide an overview here for those who are not particularly familiar with MATLAB.

Before MATLAB processing, we have digitised the graphs for Figure 3 of Jackson *et al.* 2006 [6] using Engauge Digitizer. The `.txt` export, the `.png` image, and the `.dig` save file for each of the three sub figures are provided in the directory `NewData`.

First, we open the current image for each provided file in turn, as the current image should be dominated by any feedback effects. We then perform a Fourier transform on each fast-scan line and take its modulus to get a power spectrum for each line. We then take the median of the power spectrum for each frequency to provide an averaged power spectrum. The reason for taking a median is that as the feedback is unstable it is possible to have spurious line with a particularly high or low power at a particular frequency. The median allows us to remove the effect of spurious outliers. Figure S2 shows a plot of this resulting averaged power spectrum plotted on a linear scale. The broad peak from the feedback instabilities is clear. Just below the legend horizontal lines are plotted showing the frequencies and errorbars for the "noise" from Jackson *et al.* 2006 [6].

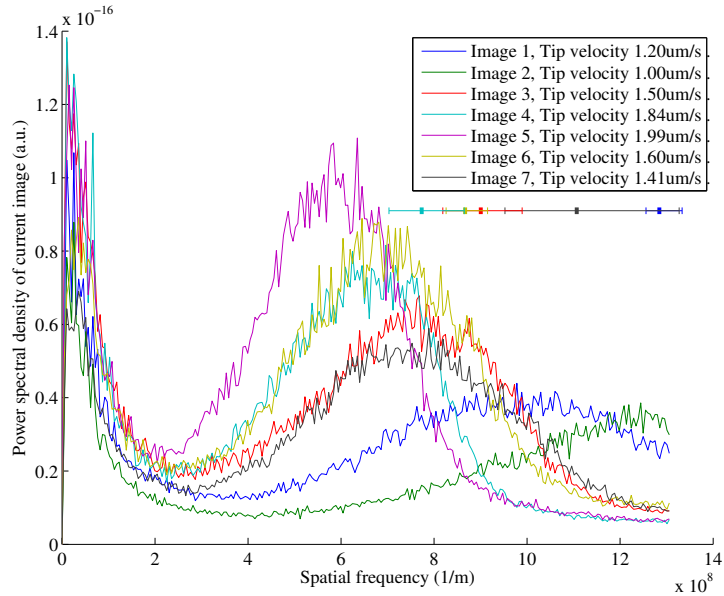

**Figure S2.** Median power spectra for the archived data provided fro Jackson *et al.* 2006 [6]. Horizontal lines below legend represent range of values for "noise" from Jackson *et al.* 2006.

To extract the full width at half maximum we are aware that the frequencies for some images are greater than the Nyquist frequency. We therefore calculate instead the half width at half maximum from the low frequency side of the curve. This is then used as the upper and lower error bar. To do this we first truncate the data to remove the low frequency components, then apply a 31-point boxcar average (31 points were chosen to produce a smooth curve without significantly affecting the shape of the overall curve) to the data and then locate the global maximum. From the global maximum we search for the first point lower than half the maximum value on the low frequency side. The smoothed data and located maxima and half maxima are shown in Figure S3

# 6 Pixelated data (Figure 5 and 6)

The data from Yu and Stellacci were not provided in the public archive of data but were sent (by Prof. Yu) to one of the authors (PJM) via private e-mail communication. Unfortunately, however, the parameter file for the experiments was not included in the data sent by Yu to PJM. The parameter file for Omicron experiments contains all meta-data for each image, including the width, height, scaling for the $z$ axis, tip speed, and all other imaging parameters. These files could not be opened with SPIW due to the lack of this file. A different software package, Gwyddion, can open them (assuming a square image) and produces images with an arbitrary $z$ scale and unknown dimensions. ASCII exports of this data, from Gwyddion, was used for processing this data in MATLAB. The images are known to be 80 nm in width from provided `.tiff` files. All raw and exported data is provided in the directory `Small`.

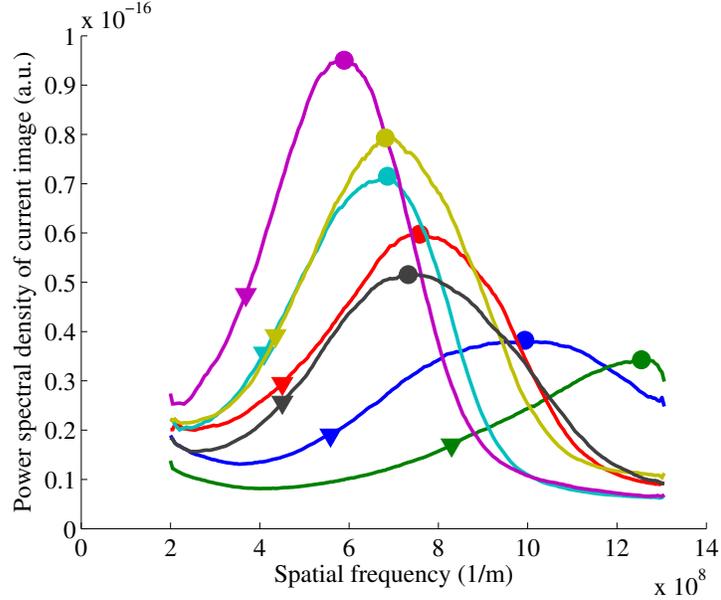

**Figure S3.** Truncated median power spectra after a 31-point boxcar average. Located maxima and half maxima are shown with circles and triangles respectively. Colours are consistent with Figure S2.

The method used to process Figures 5 and 6 are detailed in the main text. The code used is provided (in a commented form) in the files: `Small/AddIms.m` and `Small/MakeFigure.m` respectively. This relies on a provided function:
`Small/compare_trace_small_2012.m`.

In this directory there is also an interactive tool (`Small/InvestigateGUI.m`) to simultaneously view the same section of trace and retrace image for any of the scans provided by Yu and Stellacci. This tool shows both images before and after interpolation. The zoom size, position, and contrast can be modified using a GUI interface.

# 7 Figure 7

Figure 7 is entirely produced from figures in Ong *et al.* [7] with no further processing.

# 8 Persistence of tip-induced features (Figure 8)

Figure 8 was produced from the dataset described above for Figure 3 (Section 4). No image processing except first order plane flattening is applied to these images. The code used to export the images is provided in the file `NewData/BecomeJanus.m`.

# 9 Clustering of randomly positioned particles (Figure 9)

The top left digitised image was produced in GIMP by overlaying the features used in other images onto Figure 2(b) of Ong *et al.* [7]. A higher resolution copy of this overlay is provided in Figure S4 with the partial transparency of our figure showing the blue circles from the original figure. The other 8 panels were produced by randomly placing features (rejecting any that overlapped). The code used to generate these is provided in `ACSNano_n_Langmuir2013/Random_speckle.m`

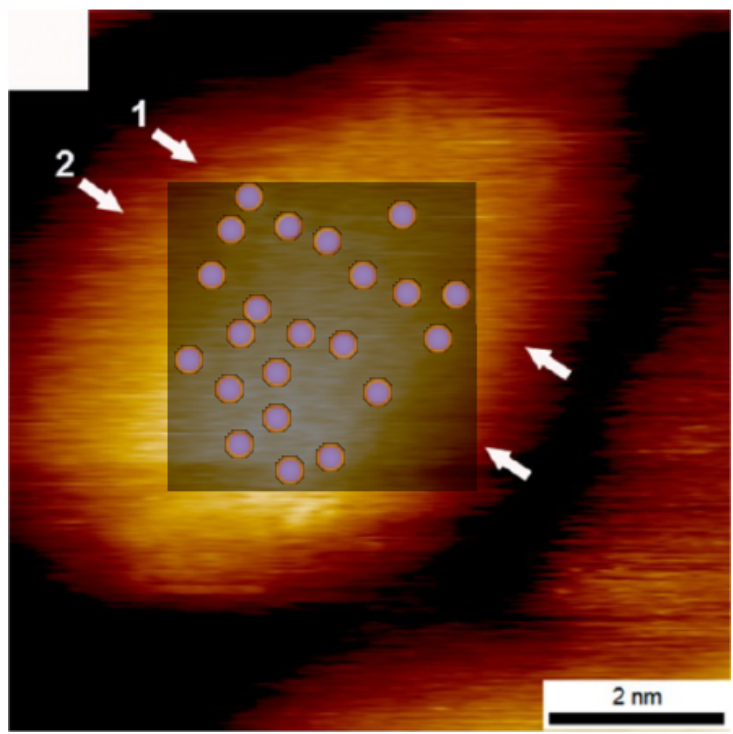

**Figure S4.** The top left panel of Figure 7 from the main text, semi-transparent and overlaid on Figure 2(b) of Ong *et al.* [7]

# 10 Features in 1D PSDs (Figure 10)

The code used to generate Figure 10 is provided in the file:
`ACSNano_n_Langmuir2013/PSD_stripy.m`. This file uses the function `Speckle.m` (also used for Figure 9) in the same directory and also functions in the `Feedback` directory used to create the surface for the feedback simulation.

# 11 Further detail on fitting in Biscarini *et al.*

The approach for extracting spatial frequencies in Biscarini *et al.* [3] involves a seven parameter fit to a function which assumes exactly two plateaus (and associated decays) added to $1/f$ noise. Non-linear fitting using high numbers of parameters, and especially those containing power laws [8, 9], are very sensitive to initial conditions and can give results which vary strongly depending on the choice of starting values.

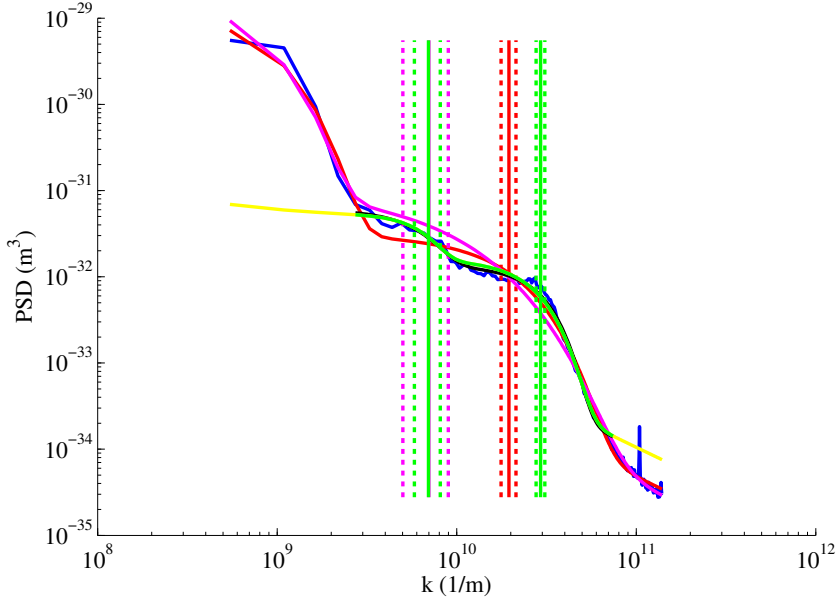

**Figure S5. Fitting a 1D PSD from Biscarini** *et al.* **[3].** Blue curve shows a 1D PSD from Figure 3a of Biscarini *et al.*. Red and magenta curves show fits to this curve using the equation adopted for fitting by Biscarini and co-workers, with different seeded (i.e. initial) values. Vertical lines of the same colour show the characteristic frequency of the second shoulder returned from the fit. The black curve (mostly obscured by the green curve) is the fit presented in the supplementary information of Biscarini *et al.*, fitted to a reduced data range. The green is our attempt to replicate the black fit, with vertical lines showing the characteristic frequencies of the first and second shoulder. As the first shoulder was discarded, Biscarini *et al.* use the first of these frequencies to define the second shoulder, even though their model expects only two shoulders and the next shoulder dominates. The yellow curve is an extension of the green to show that the extended fit poorly fits the remaining excluded data.

Figure S5 shows two functions (magenta and red) fitted to the blue data from Figure 3a) of Biscarini *et al.*. Vertical lines of the same colour show the position of the extracted characteristic frequency with 95% confidence bounds marked by dashed lines. Even more misleading is the analysis of this curve in the original paper where only a small range was fitted, excluding the first plateau and some of the final noise [10]. Our attempt to recreate this fit (green) shows two characteristic frequencies marked by vertical lines. As the initial plateau has been discarded, Biscarini *et al.* take the first of these frequencies as being their second shoulder frequency despite the second fitted shoulder clearly dominating and their model only predicting two rather than three shoulders! This fitted frequency agrees with two further fitted frequencies for other 1D PSDs which is used to suggest consistent frequencies between images, while a fourth which does not agree is discarded as an outlier.

It is therefore clear that the fitting is far from robust, and that sections of data were discarded to improve the consistency without this exclusion being mentioned in the text. We further note that all fits, including the green curve, were computed in MATLAB and gave warnings for poor convergence.

The above analysis was performed with Igor files provided by Fabio Biscarini (`.pxp` files) which are included in the directory `ACSNano_n_Langmuir2013`. The data named `lnk11` and `lnPSD11` and the corresponding fit have been imported into MATLAB and saved as the `.mat` file: `DataFromIgor.mat`. The function `FittingForFigure.m` fits for the full data using the equation from Biscarini *et al* [3] for two seeded parameter sets using the MATLAB function `lsqcurvefit` for non-linear fitting. A further fit (as explained in the main text) is produced for a decreased range to attempt to match the fitting of Biscarini *et al.*.

# 12 NMR spectroscopy

## 12.1 Further details on 1D spectroscopy of broad aryl peak

### 12.1.1 Further details for Janus nanoparticles

In the main text we describe problems with the linear model for the change in chemical shift presented in Equation 1 of Liu *et. al.* [2] for randomly mixed nanoparticles. The chemical shifts arise from the effects of localised magnetic fields generated in the aromatic rings of the DPT. For the Janus nanoparticles there is another fundamental problem with their approach. The authors assume that the position of the peak maximum, $F$, can be calculated as a superposition of two discrete chemical shifts, one originating from ligands in a bulk-like environment, and one coming from the interface between the two regions of the Janus particle.

This would be correct for two sharp, well separated lines that coalesce because of a fast mixing of the two components [11]. However, the two components clearly do not mix, since each ligand is fixed on the surface of the nanoparticles. In addition, the two lines are broadened due to an incomplete motional averaging of either anisotropic (orientation-dependent) interactions, such as dipole-dipole couplings or chemical shift anisotropies, or the presence of magnetic field gradients at the interface between the nanoparticles core and the organic shell due to different susceptibilities of the two. The two broad lines overlap significantly and the linewidths of the two regions are not necessarily identical (this would depend on the motion of the ligands in each domain as well as their packing). The qualitative trend of the expected peak position as a function of composition $x_A$ should still be correctly described by the equation put forward by Liu *et al.*, but a correct quantitative description of the interface layer thickness $t$ would only be coincidental. A quantitative estimate of $t$ from an NMR data set may be possible by decomposing the line into a bulk component and a interface component, provided that the two lineshapes do not change as a function of $x_A$.

Aside from these fundamental limitations, the "rigorous" model presented by Liu *et. al.* [2] converges incorrectly at both concentration limits and leads to a clear trend in residuals. This comes from their treatment of the interaction band $t$ being defined along the cross section of the particle rather than as a band of thickness $\tilde{t}$ on the surface of the particle (see Figure S6). This clearly has a strong effect at low or high concentrations where $t$ should decrease in size, but remains constant in their treatment. Instead, the interaction area should be

$$A_{\text{int}} = 2\pi r_c \tilde{t} = 2\pi r \sin(\theta_c)\tilde{t},$$ (1)

where $r$ is the radius of the nanoparticle, and $\theta_c$ is the opening angle of the interface layer from the centre of the nanoparticle sphere. This form of the equation is not directly useful as we need to define $\sin(\theta_c)$ in terms of the height, $h$, of the spherical cap. From basic trigonometry

$$h = r(1 - \cos(\theta_c)) \qquad \rightarrow \qquad cos(\theta_c) = 1 - \frac{h}{r},$$ (2)

and using the identity $\cos^2(\theta) + \sin^2(\theta) = 1$:

$$\left(1 - \frac{h}{r}\right)^2 + \sin^2(\theta_c) = 1$$ (3)

$$\therefore \quad \sin(\theta_c) = \sqrt{1 - \left(1 - \frac{h}{r}\right)^2} = \sqrt{\frac{h}{r}\left(2 - \frac{h}{r}\right)}.$$ (4)

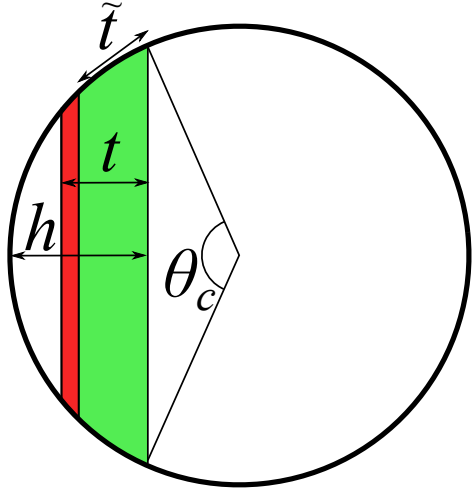

**Figure S6.** Area of the interaction stripe should be considered for an area thickness $\tilde{t}$ along the surface (green area) rather than a constant area thickness $t$ along the cross section, otherwise the interaction area is significantly over estimated (red area) at high or low concentration.

Now combining Equations 1 and 4 we get:

$$A_{\mathrm{int}} = 2\pi r \tilde{t} \sqrt{\frac{h}{r}\left(2 - \frac{h}{r}\right)} = 4\pi r \tilde{t} \sqrt{x_A(1 - x_A)}, \tag{5}$$

where $x_A = h/2r$ is the mole fraction of the thiol being detected.

In analogy to Liu *et. al.* [2], we take the equation for the area of a spherical cap to be the area taken for all molecules of this thiol,

$$A = 4\pi r^2 x_A. \tag{6}$$

Applying a constant chemical shift $B$ for of thiols outside of the interaction band, covering an area $A - A_{\mathrm{int}}$, and a chemical shift $I$ for thiols within the interaction band , we arrive at a mean chemical shift of

$$F = \frac{(A - A_{\mathrm{int}})B + A_{\mathrm{int}}I}{A} \tag{7}$$

$$= B + \frac{\tilde{t}\sqrt{x_A(1 - x_A)}(I - B)}{r x_A}. \tag{8}$$

This model, as mentioned in the main text, is not truly rigorous either because it is only valid for areas $A$ where

$$A \geq A_{\mathrm{int}}. \tag{9}$$

Notice that in the current definition, $A_{\mathrm{int}}$ only represents the fraction of the area within the interaction stripe that is covered by thiol A.

The data from Figure 3e in Liu *et. al.* [2], has been digitised using Engauge Digitizer (the digitised graphs are provided in the `NMR` folder of the supplementary information). Fitting the above model to the data using a least squares fitting algorithm in MATLAB, we arrive at a functional form of

$$F = 6.993 + \frac{.1464 \times \sqrt{x_A(1 - x_A)}}{x_A} \tag{10}$$

where each data point is weighted by its vertical errorbar. See Figure S7 to compare the quality of the fit for the two models. Our model has an $R^2$ of .997 with no data points excluded, compared to the model of Liu *et. al.* with $R^2$ of .976 after arbitrarily excluding a point. This model is derived in full not because it has a direct impact on the evidence for striped morphologies for thiol capping layers, but instead to demonstrate careless data analysis and modelling that could easily be improved using the simple model described above.

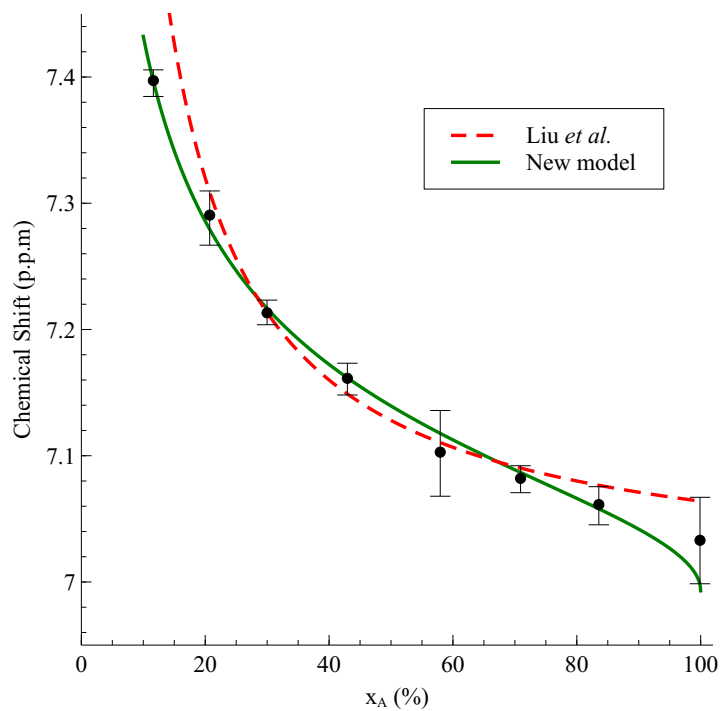

**Figure S7.** Digitised data for aryl peak chemical shift of Janus naoparticles from Liu *et. al.* [2]. The lines represent fits to the model from Liu *et. al.* (dashed) and our new model (solid).

### 12.1.2 Further details for patchy/striped nanoparticles

Liu *et. al.* [2] claim that the combination of their results they present can only be explained by the formation of stripe-like domains in the patchy nanoparticles. In the main text we suggested a different model that could also explain Liu *et al.*'s data, which we explain in more detail here. In this model the NMR data could also be caused by DPT molecules forming circular patches of well-defined size, with the phenyl rings stacked against each other. If the DPT/DDT ratio determines the number of patches, but not their size, the chemical shift of the NMR signal caused by the DPT would not significantly change over a relatively large concentration range. Essentially, from the point where enough DPT is present to form patches to the point where patches get so close to each other that they start interacting, the chemical shift would be constant. If the patches wwew circular or near-circular, and were big enough to accommodate a single DPT molecule in their centre, such a central DPT molecule may be able to rotate reasonably freely, which would cause a significant motional narrowing of its linewidth. However, rotation about the third axis would still be hindered, as this would require the whole nanoparticle to rotate, and hence some residual broadening is expected. We also note that ff the wall of the patches is caused by stacked aryl rings, the central molecule would be in the plane of these rings, hence not experiencing an upfield shift (towards lower ppm values). In addition, the space on the nanoparticle's surface that is available to this molecule would be larger than the space available to DPT molecules that form the patch wall. Hence a larger reactivity could be expected. The narrow peak that is apparent in the broad aryl line could be explained by such an arrangement.

Moreover, the analysis of the ratios of the peaks presented in Table 2 would suggest that the patch wall consists of 16 to 26 DPT molecules, which would be sufficient to accommodate a single DPT molecule on the inside. Furthermore, due to the star-shaped arrangement of the aryl rings in the patch wall, the interface between DPT and DDT molecules would be fairly substantial (more substantial than in the case of a striped arrangement). Hence cross peaks in a 2D NOESY experiment would be expected, and their amplitude would even be larger than for a striped arrangement.

Although the current data cannot conclusively distinguish between such a patchy and a striped geometry of the ligand molecules, it is consistent with both interpretations. Hence, the NMR data presented does not conclusively show the presence (or absence) of stripe-like domains.

## 12.2   Discussion of narrow aryl peak

Liu *et. al.* [2] performed diffusion-ordered spectroscopy (DOSY) to measure the mobility of the broad aryl peak in comparison with the narrow aryl peak. The corresponding spectra are shown in the Supplement (Figure S14). Panels a and b show basically identical diffusion constants for the ligands that are causing the two peaks. This is to be expected, since the DOSY experiment essentially provides the mobility of the nanoparticles. Whether a ligand is covalently bound or non-covalently adsorbed on the surface cannot be assessed with this technique, since the distance travelled by an adsorbed molecule on the surface would be too small to induce an echo decay in a DOSY experiment: an adsorbed molecule would only travel a distance of $\pm 2$ nm compared with the center of mass of a nanoparticle with a diameter of 4 nm, which is about three orders of magnitude too small a distance to be picked up with the employed gradient strengths [12]. That the molecule giving rise to the narrow peak is sticking to the surface of the nanoparticles in some way had already been established chemically, hence the DOSY data does not provide any additional information.

In addition, the analysis of the DOSY data is lacking an error estimate for the diffusion constants. The fits in Figure S14 a and b, using the Stejskal-Tanner equation, are not perfect, but the residuals are not very pronounced. Hence the assumption of a single diffusion constant appears to be a practical working model. However, the data for the mixture of nanoparticles and free ligand in panels c and d show very clear trends of the residuals. Fitting a single diffusion constant is misleading, and the obtained data does not represent a physical picture of the motion in the sample. An interpretation of the DOSY data in a two-dimensional way, where the chemical shift is correlated with a distribution of diffusion constants [13], would be a much more suitable method in this case. However, the information that could be obtained with respect to the ligand shell of the nanoparticles would still be very limited.

To demonstrate the increased reactivity of the molecules that cause the narrow aryl line, an experiment is presented where ligands were exchanged using aminoanthracene. The NMR spectra of the nanoparticles before and after this exchange were then compared (Figure 6). The two spectra were approximately scaled that the broad peaks had a comparable amplitude. However, the signal-to-noise ratio of the spectrum after the exchange is about an order of magnitude lower than before the exchange, and the lineshape of the broad line differs significantly between the two spectra. No experimental details were provided that would explain these differences. Assuming that the amount of sample was the same in both cases and that the spectra were recorded with an identical protocol, these data would imply that the aminoanthracene caused much more significant changes to the nanoparticles shell than just simply replacing the mobile aryl molecules, as suggested in the paper. A quantitative statement about the reactivity of the aryl molecules that cause the narrow peak in the original nanoparticles would require the spectrum after the exchange to be fully interpreted, including the sudden appearance of very narrow peaks and the change of the shape of the broad line. In the current form, the two spectra are too different to conclusively show any of the claims made by the authors regarding the ligand reactivity.

## 12.3  Discussion of 2D NOESY data

Liu *et. al.* [2] also use 2D nuclear Overhauser exchange spectroscopy (NOESY) to analyse the ligand structure of gold nanoparticles. The authors base their arguments on the occurrence of cross-peaks between the alkyl peaks (occurring at a chemical shift of about 1 ppm) and the aryl peaks (with a chemical shift of about 7 ppm) in the NOESY data. Again the authors are selective in the information they analyse. Only the presence of a cross-peak or lack thereof is considered. However, the notable change in shape of the diagonal peaks is ignored. Such changes in diagonal peaks shapes can indicate a change in mobility.

When introducing the NOESY technique, the authors only state the dependence of cross-peaks on the distance between ligands. However, the cross-peak intensity is a product of the spectral density of the ligand motion at particular frequencies and, as correctly stated, the inverse of the distance between ligands to the sixth power [11]. As a consequence, a quantitative analysis of NOESY data necessarily requires to take the distance as well as the mobility into account. Certain ligands may show no cross-peak, even if they would be within the stated 0.4 nm from each other: the NOESY cross-peaks change their sign at a particular correlation time constant of the ligand motion, and when they do, the cross-peaks disappear.

Even if cross-peaks are not cancelled completely, a change in mobility, as caused by a size variation of the nanoparticles, can cause significant amplitude variation. If the cross-peak amplitude is close to the resolution limit, as appears to be the case in the presented data, a change of the diameter of the nanoparticles by a factor two, which is the size difference between the Janus nanoparticles that do not show cross-peaks and the other nanoparticles that do show cross peaks, could be the primary reason for the apparent absence of cross-peaks. Therefore a general statement that Janus particles can be distinguished by the absence of NOESY cross-peaks cannot be confirmed by looking at a single particle size only, especially since the other types of nanoparticles, which did show cross-peaks, had a significantly different size. Since this is the only information that was extracted from the NOESY data, these spectra do not seem to be useful for any conclusive statement without a more in-depth quantitative analysis. Such an analysis would have to consider not only the presence of cross-peaks, but also their amplitude and sign, as well as the mass and type of the nanoparticles.

Another shortcoming of the NOESY data is the apparently inconsistent use of contour levels between the spectra. This prevents a quantitative comparison of the cross-peak amplitudes from the data and an estimate of their lower limit. After all, even in the Janus nanoparticles there exists an interface between the two ligands that is not negligible for nanoparticles with a diameter of 2-3 nm, provided that the stated value of 0.4 nm for the layer thickness is of the right order of magnitude. Hence, although weaker, a cross-peak should become visible at some point for experimental conditions where cross-peak are not cancelled.

## 12.4  Section summary

In summary, the presented NMR data is not suitable to draw any conclusion about the structure of patchy/striped nanoparticles. This is not to say that NMR is an unsuitable tool for this kind of study, but the presented data and analysis do not support the conclusions that were drawn. The one exception is the authenticity and reproducibility of the narrow peak in the DPT-DDT nanoparticles, which was established chemically.

We would like to conclude by drawing an analogy, from an NMR point of view, between proteins and nanoparticles that are surface-coated with organic molecules. Despite their very different physical and chemical properties and different applications, the two may not be that different from an NMR point of view. Their size, shape and mass can be comparable, and, hence, both may show similar mobility in a solvent.

While many nanoparticles cores are not directly accessible to study by NMR, an organic coating would be. If such a coating consists, as in the present case, of small molecules that can be easily distinguished by their chemical shift, many techniques that were developed for protein NMR may be applicable for the study of the surface coating of nanoparticles. Therefore, the choice of NMR by Liu *et. al.* [2] to probe the surface structure of nanoparticless is plausible, and NMR may in fact be able to answer many questions about the structure of organic molecules on the surface of nanoparticles in great detail.

NMR spectroscopy would not, however, be able to replace any of the imaging techniques, such as STM or TEM, for the study of nanoparticles. The information content of an NMR experiment is very different and in fact completely complementary to most of the techniques currently used routinely to study nanoparticles. It would appear to be worthwhile to apply the huge pool of methods and the experience in performing NMR experiments with large, low mobility proteins for the study of surface-coated nanoparticles. Collaborations between nanoparticle specialists and protein NMR groups, therefore, may prove to be very fruitful.